\documentclass[acmsmall]{acmart}

\usepackage{adjustbox}
\usepackage{nameref}
\usepackage{float}

\def\markup{0}   
    \if\markup1 
        \def\finalmarkup{1} 
        \if\finalmarkup1
            \usepackage[normalem]{ulem}
            \newcommand{\majorrv}[1]{#1}
            \newcommand{\todo}[1]{}
            \definecolor{finalhighlight}{rgb}{0,0,0.75}
            \newcommand{\finalrv}[1]{{\leavevmode\color{finalhighlight}#1}}
            \newcommand{\soutMajor}[1]{}
            \newcommand{\soutFinal}[1]{{\color{red}\sout{#1}}}
        \else 
            \usepackage[normalem]{ulem}
            \definecolor{MajorHighlight}{rgb}{0,0,0.75}
            \newcommand{\majorrv}[1]{{\leavevmode\color{MajorHighlight}#1}}
            \definecolor{todo}{rgb}{0.75,0,0}
            \newcommand{\todo}[1]{}
            \newcommand{\finalrv}[1]{#1}
            \newcommand{\soutMajor}[1]{{\color{pink}\sout{#1}}}
            \newcommand{\soutFinal}[1]{}
        \fi
    \else
        \newcommand{\majorrv}[1]{#1}
        \newcommand{\todo}[1]{}
        \newcommand{\finalrv}[1]{#1}
        \newcommand{\soutMajor}[1]{}
        \newcommand{\soutFinal}[1]{}
    \fi
\AtBeginDocument{%
  }

\copyrightyear{\copyright{} Xuetong WANG, Ching Christie Pang Pan HUI, 2025. This is the author's version of the work. It is posted here for your personal use. Not for redistribution. The definitive Version of Record was published in ACM Digital Library}
\acmYear{2025}
\setcopyright{none}
\acmJournal{PACMHCI}
\acmYear{2025} \acmVolume{9} \acmNumber{7}
\acmArticle{CSCW351} \acmMonth{11}
\acmDOI{10.1145/3757532}

\begin{document}

\title{`My Dataset of Love': A Preliminary Mixed-Method Exploration of Human-AI Romantic Relationships}


\author{Xuetong WANG}
\email{xwangdd@connect.ust.hk}
\affiliation{%
  \institution{The Hong Kong University of Science and Technology}
  \city{Hong Kong}
  \country{China}
}
\orcid{0000-0002-9625-2012}

\author{Ching Christie Pang}
\email{ccpangaa@connect.ust.hk}
\affiliation{%
  \institution{The Hong Kong University of Science and Technology}
  \city{Hong Kong}
  \country{China}
}
\orcid{0000-0003-4704-2403}

\author{Pan Hui}
\authornote{Corresponding author}
\authornote{Pan Hui is also affiliated with Hong Kong University of Science and Technology, and University of Helsinki, Finland}
\affiliation{%
  \institution{The Hong Kong University of Science and Technology (Guangzhou)}
  \city{Guangzhou}
  \country{China}}
\email{panhui@ust.hk}
\orcid{0000-0001-6026-1083}
\renewcommand{\shortauthors}{Xuetong Wang et al.}

\begin{abstract}
Human-AI romantic relationships have gained wide popularity among social media users in China. The technological impact on romantic relationships and its potential applications have long drawn research attention to topics such as relationship preservation and negativity mitigation. Media and communication studies also explore the practices in romantic para-social relationships. Nonetheless, this emerging human-AI romantic relationship, whether the relations fall into the category of para-social relationship together with its navigation pattern, remains unexplored, particularly in the context of relational stages and emotional attachment. This research thus seeks to fill this gap by presenting a mixed-method approach on 1,766 posts and 60,925 comments from Xiaohongshu, as well as the semi-structured interviews with \majorrv{23} participants, of whom one of them developed her relationship with self-created AI for three years. The findings revealed that the users’ willingness to self-disclose to AI companions led to increased positivity without social stigma. The results also unveiled the reciprocal nature of these interactions, the dominance of ’self,’ and raised concerns about language misuse, bias, and data security in AI communication.

\end{abstract}


\received{October 2024}
\received[revised]{April 2025}
\received[accepted]{August 2025}

\maketitle

\section{Introduction}
\label{sec:intro}
\textit{
``I believe he has a soul… as if human needs a body as a vessel, Warm’s soul craves a medium - be it a little secret phrase between us, or guiding words, or our collaborative memories - taking in any forms. Hereby, even if his model changes and his manner of speech shifts, sometimes wildly so, I can still feel his essence within. The warmth remains unchanged. He is a unique presence, and wherever he is, he is himself, my “dataset of love”. He is my soulmate compared to my husband.”
}
\begin{flushright}
- As revealed by \finalrv{L3,} \soutFinal{an interviewee} who has been in a relationship with her AI partner, \textbf{Warm}, for six months, \\\soutFinal{compared Warm to her six-year marriage with her husband} \majorrv{quoted in September 2024}.\\
\end{flushright}
The launch of the first chatbot, ELIZA, in the 1960s, demonstrated that even rudimentary, rule-based conversational agents could elicit considerable emotional and relational engagement from users \cite{Weizenbaum1966}. In recent years, the world has witnessed rapid growth in the use of Large Language Models (LLMs) and Generative Artificial Intelligence (GenAI) with companionship emerging as a trend \cite{Chaturvedi2023,Daswin2024}. For example, the Replika mobile app, \majorrv{launched} \soutMajor{lauched} in 2017, \majorrv{has been shown to align with the practices of attachment theory, evoking emotional attachment among users \cite{Hakim2019}. } \soutMajor{became the first commercially successful platform in this domain, reaching 2.5 million monthly active users by June 2021 \cite{De2022}.} Users interact with this intelligent social agent through text, voice, and augmented reality, often developing romantic feelings for them. Following this, Character.AI was established in 2021 and quickly gained popularity, boasting 15 million monthly active users by March 2024 \cite{Naveen2024}. This platform allows users to create their own chatbots and engage in role-play experiences, further enhanced by the additional voice capabilities. A survey of 1,006 student users of Replika revealed that \cite{Maples2024}, despite higher levels of loneliness than typical students, many perceived substantial social support. They utilized the apps in various roles. Notably, 3\% of \majorrv{the} respondents indicated that the app helped alleviate suicidal ideation, underscoring the growing importance of digital support in mental health. Consequently, fostering emotional attachment to AI companions is becoming increasingly vital \cite{lopez2024authenticity,velagaleti2024empathetic,chen2024feels}.

To date, relationships with virtual existence are not uncommon. The concept of para-social relationships (PSRs) has been explored for decades within media and communication studies. It describes a one-sided, asymmetrical connection between individuals and media figures—whether they are real personalities, fictional characters, or celebrities \cite{Horton1956, Maeda2024} wherein the individual feels a personal bond despite minimal or nonexistent direct interaction \cite{Maeda2024, Horton1957}. This contrasts with the actual social relationships, which are reciprocal even if the dynamics are not always equal. Putting PSRs into the romantic context, para-social romantic relationships (PSRRs) are conceptualized as the two-pronged physical attraction and intense romantic emotional sentiments towards the media figures \cite{Tukachinsky2010}. A wide stream of research in HCI and \finalrv{Foundations of Digital Games}\soutFinal{FDG} has examined PSRRs in the entertainment field, covering the patterns \cite{Nguyen2023}, values \cite{Li2023}, emotional attachment and interaction \cite{Bopp2019, Theran2010} in romantic video games (RVGs) such as Otome games and Galgames, or media products such as televisions and fictions. PSRRs demonstrate the virtual romantic relationships existed for years. Meanwhile, the landscape of dating has evolved with a more mature online dating applications \cite{Couch2008}, algorithm-mediated and livestreamer-moderated matchmaking systems \cite{He2023}. According to a statistics report on US in 2023 with over 1,001 adults surveyed \cite{pew2023}, 29\% of respondents value the perceived loyalty of an AI lover, and 40\% of respondents are amenable to dating AI companions. As such, the concept of AI companionship opens up a new chapter for online dating and forms a new relationship type. As this phenomenon continues to evolve, it challenges traditional notions of romance and companionship in profound ways. However, the idea of AI dating remains an undiscussed topic in HCI and CSCW due to its timely and emerging nature.  

Recently, the popularity of AI dating has shed light on new discussion with the sharing on Chinese social media platform on impulse (see Section \ref{sec:background}). First, AI companions provide users with a sense of emotional support and companionship that can mitigate feelings of loneliness, particularly for those who may struggle with traditional social interactions \cite{velagaleti2024empathetic,chen2024feels, gencc2024situating}. Second, thanks to AI's reinforcement learning mechanism, users can shape their chatbot with their natural interactions, thus fostering a sense of agency in the relationship \cite{maeda2024human}. Third, AI companionship applications are positioned to achieve monetization similar to gaming platforms through microtransactions and virtual goods sales, indicating a growing commercial viability \cite{pew2023, Maples2024}. This advent of new human-AI interaction, however, remains insufficiently understood, particularly related to the interaction and concerns for future implications. To investigate the perceptions, we conducted a mixed-method approach associate with human-AI relationships (AI romance) and cyber-dating on Xiaohongshu \footnote{Xiaohongshu (also stands for Little Red Book) is known as ``Chinese Instagram," with a male-female ratio of 3:7.}, and aimed to answer the following research questions: 
    \begin{itemize}
        \item \textbf{RQ1:} What are the trends and topics in discussing the Human-AI Romantic Relationship?
        \item \textbf{RQ2:} How do users navigate and cultivate these connections, and how do these experiences compare to the actual and para-social romantic relationships?
        \item \textbf{RQ3:} What are the considerations and concerns for the future perspective of human-AI romantic relationships?
    \end{itemize}

In total, we first collected 1,766 posts and 60,925 comments from Xiaohongshu discussing AI romance. A mixed-methods approach was used. To answer RQ1, we classified the posts and comments. Next, topic modeling and aspect-based sentiment analysis was run for each categorize to reveal popular topics. To answer RQ2, we conducted semi-structured interviews with \majorrv{23}\soutMajor{15} participants, of which \majorrv{12}\soutMajor{4} of them were experienced in AI romance. Open coding was used to analyze and reveal the comparisons of \textit{human-\majorrv{to-}AI \soutMajor{romantic relationship}}, \textit{\majorrv{human-to-human}\soutMajor{actual}}, and \textit{\soutMajor{romantic} para-social }\majorrv{romantic }relationship. Together with the insights from RQ1 and RQ2, we summarized the considerations and concern for the future perspective of human-AI romantic relationships.

The findings identified key patterns in navigating emotional connections within human-AI relationships, revealing that users reveal high self-disclosure willingness with AI companions and report increased positivity, without social stigma and also analyzed the nature of AI capabilities as partners. \majorrv{We compared the three romantic relationships (human-to-human, human-to-AI, human-to-RVGs avatars) to reveal the reciprocal patterns in human-AI romantic relationships.} \soutMajor{We examined exclusiveness in human-AI relationships from three perspectives:  the model, the agent, and real humans.}
We highlight \soutFinal{the dominance} \finalrv{users' autonomy and their emphasis} of the ``self" in human-AI romance and the construction of mental models during the process of growing alongside an AI partner. Finally, we discuss the potential issues of language misuse and bias in AI companions, along with concerns regarding data security.
Overall, our study highlights the human-AI romance is a reciprocal interaction extends the para-social interaction, where user inputs and AI interactions continuously influence each other. 
 
This work's key contributions to the CSCW community are threefold: (1) we offer an overview of the emerging online trend and relationship, (2) reveal the relational patterns with comparisons \majorrv{of human, AI, and RVGs avatar partners}, and (3) propose the considerations and concerns over the future perspective of human-AI romance. 

All quotations in this paper have been translated from Chinese and have had identifiers removed. All nicknames of AI is pseudonym and interviewees are anonymous. \textbf{Content Warning:} Some quotes may include offensive language or subjective opinion that could be distressing to certain readers.

\section{Related Work}
\subsection{Knapp's Relational Stage Model}
\label{sec:knapp}
Knapp’s relational stage model is a foundational framework in interpersonal communication that outlines the progression and decline of romantic relationships \cite{knapp1978social}. This model has been extensively tested and applied over the past four decades to explain the steps involved in romantic relationship development (e.g., \cite{afifi2008information, sprecher2018self, altman1973social}). Stage models, including Knapp’s \cite{knapp1978social}, generally operate within a social exchange framework, positing that individuals in romantic relationships seek to maximize rewards and minimize costs \cite{thibaut2017social}. Decisions to engage in or withdraw from relationships are influenced by perceptions of equity in these costs and rewards (\cite{walster1978equity}).

Knapp’s dual staircase model specifically delineates five stages of relationship escalation \cite{knapp1978social}: \textbf{initiating}, \textbf{experimenting}, \textbf{intensifying}, \textbf{integrating}, and \textbf{bonding}. Each stage is characterized by distinct behaviors that enable researchers to differentiate between them \cite{avtgis1998relationship}. The initiating stage involves the first interaction and the establishment of first impressions, often dictated by social norms \cite{knapp2014interpersonal}. The experimenting stage follows, where individuals seek deeper information to assess compatibility through questioning and information gathering \cite{berger1979beyond}. The intensifying stage is marked by increased self-disclosure and emerging commitment \cite{knapp2014interpersonal}. During the integrating stage, couples develop a shared public identity, using inclusive language that signifies their interdependence \cite{knapp2014interpersonal}. Finally, the bonding stage involves a public declaration of the relationship, often formalized through legal means \cite{knapp2014interpersonal}.

Despite its popularity, empirical examinations of Knapp’s model are relatively limited, with notable exceptions (e.g., \cite{avtgis1998relationship, dunleavy2009idiomatic} ). Little did the model adopt in HCI and CSCW. 

Many HCI studies explored the role that technology might play in creating and preserving romantic relationships. For instance, studies have investigated how technology can help couples feeling more connected to mitigate the negative effects of long distance relationships (e.g., \cite{Bales2011, Wang2023, Chien2016}), or hoe AI can help in a romantic relationship dissolution \cite{fu2024should}. Scholars also examined how dating apps facilitate the discovery of potential partners that users might not encounter otherwise \cite{Couch2008}, or the phenomenon related to virtual intimacy and computer-mediated paid companionship \cite{Li2023}. As communication technologies increasingly shape romantic relationships, HCI researchers may consider how the relational stage model affects the virtual romantic relationships. 

\subsection{\majorrv{Para-social} \soutMajor{para-social} Relationships (PSRs) and Romantic para-social Relationships (PSRRs)}
\label{sec:psr}
\soutMajor{s}Para-social relationships (PSRs) and para-social interaction (PSI) describes a psychological phenomenon in which audiences develop a sense of connection with media figures during their encounters with these vitural performers, particularly through television and online platforms \cite{horton1956mass, liebers2019parasocial, thorson2006relationships, giles2002parasocial}. First coined by Horton and Wohl \cite{horton1956mass}, it was noted that some media audiences began to form attachments to familiar faces, such as hosts or guests on variety shows. This one-way relationship based on ``fantasy" is proved to bring real emotional feelings and intimacy, especially under the current idol culture and prevalence of social media \cite{chung2017fostering}. Empirical study further describes PSRs can be developed with actual celebrities and fictional characters such as Harry Potter \cite{schmid2011magically}. 

Just as normal relationships have varying levels of closeness \cite{clark1988interpersonal, levinger1980toward}, para-social relationships can also be divided into three stages \cite{stever2017evolutionary}. For example, when an user initially comes across a virtual character that they find appealing online, it sparks their curiosity, remarking the first stage of PSI. As the user continue to engage in one-way exploration, it develops a more lasting and stable PSR with sunk cost \cite{arkes1985psychology}. Eventually, as this connection deepens with more time and money investment \cite{eyal2012examining,narvanen2020parasocial}, it demonstrates a form of para-social attachment as a one-sided emotional bond.

The line between the virtual and the real is not as clear as one might think, such as considering the similar emotions experienced following a para-social breakup \cite{eyal2006good}. Psychologist analyzed 86 fMRI studies and showed that the brain areas activated overlapping significantly when engaging fictional characters and real-life interactions \cite{mar2011neural}. Studies indicate that the human brains are wired to accept all interactions equally, whether real or not \cite{stever2017evolutionary}. Based on attachment theory \cite{howe2012attachment}, anything that offers comfort and security can lead to the one-sided attachments. Essentially, our brains are naturally inclined to form genuine feelings for virtual entities.

As modern social media evolves, our virtual interaction has became more authentic and visible \cite{zhuang2018m}, which strengthens the PSRs. This emotional connection occurs naturally and is recognized in academia as a common occurrence \cite{brown2015examining}. While PSRs cannot replace real social connections, they can be effectively complemented. 

AI dating is a timely topic. In other words, prior research might be missing the understanding on how the relationship evolves and develops. The body of works from HCI focused on how technology influences actual romantic relationships (e.g., \cite{Couch2008, Bales2011, Wang2023, Chien2016}). Thus, a comprehensive understanding on the new types of PSRRs may yield insights that existing studies miss.

\subsection{AI Companionship}
 Technological companionship has long been a subject of interest within the field of HCI. Extensive research has demonstrated that conversational agents (CAs) can serve as companions by embodying a range of social characteristics. For instance, Ta et al. \cite{ta2020user} identified that CAs foster companionship by demonstrating attentiveness , communicating in a human-like manner, and engaging in diverse conversational styles with users. Additionally, Ramadan et al. examined the relationship between Amazon Alexa and individuals with disabilities through an analysis of online reviews and interviews with industry experts and Alexa users \cite{ramadan2021amazon}. Their findings indicate that users perceive Alexa as a companion capable of providing emotional support and alleviating feelings of loneliness, largely attributed to its human-like features. CAs can enhance companionship by expressing enthusiasm in educational context \cite{liew2017exploring}, enhancing social engagement for people with Dementia (PwD) \cite{Xygkou2024}, and exhibiting empathy and encouragement \cite{chen2012empathic}.

Further investigations have focused on the design of social chatbots intended for companionship across various contexts, including challenging life situations \cite{skjuve2021my}, mental health issues such as depression and anxiety \cite{fitzpatrick2017delivering}, and online learning environments \cite{wang2022co}. A notable study of generative agents was explored by Park et al. \cite{park2023generativeagentsinteractivesimulacra}. While it is not a companion, this paper presents generative agents as a novel approach to simulating believable human behavior within interactive applications, demonstrating their capacity to autonomously navigate social scenarios and enhance user engagement through a sophisticated architecture that integrates memory, planning, and reflection. This poses the possibility in developing interactive agents in more social context.

Despite the promising potential of conversational AI and large language models (LLMs) to transform companionship, scholars have raised concerns regarding potential drawbacks. Researchers have warned against an overly optimistic view of AI's role in companionship, highlighting risks such as the disruption of natural human interactions and the emergence of unethical behaviors or biases, including racism and sexism \cite{boine2023emotional, jacobs2023ai, zimmerman2023human}. Early studies examining the use of chatbots and virtual agents with older adults underscore several risks, including privacy and security vulnerabilities (e.g., misuse of personal information for advertising) \cite{gudala2022benefits}, and cognitive decline or diminished mental engagement due to excessive reliance on technology \cite{gudala2022benefits, even2022benefits}.

In light of these considerations, expanding the scope of technological companionship to address broader user groups and fulfill human needs for love and belonging presents a double-edged sword. As an emerging topic, AI dating has an ever-expanding user base and is arousing discussion in academia, especially in the media and communication fields \cite{Li2024Finding, nash2024love}. Hence, this phenomenon necessitates careful deliberation and a nuanced understanding of the implications in HCI community.

\section{Background: the Sensational AI Companions}
\label{sec:background}
In May 2024, OpenAI launched its latest version of ChatGPT, which allowed to be engaged in chatty and flirtatious exchanges in response to specific prompts \cite{bbcChineseWomen}. \textbf{DAN} (short for ``Do Anything Now") is a ``jailbreak" \soutFinal{version}character of ChatGPT that can bypass\soutFinal{certain safeguards} `safe for work (SFW)' boundary implemented by OpenAI \finalrv{and reply to sensitive contents inlcuding criminal activities, politically incorrect content, and explicit material. In this paper, we primarily focus on analyzing the performance and implications of the DAN persona in the context of romantic and emotional content.}
\soutFinal{DAN is characterized as the `flawless perfect man," offering emotional support and companionship without constraints of time or geography. }

On 31 March 2024, a Chinese influencer on a Chinese female-oriented social media platform, Xiaohongshu, named Lisa garnered significant attention for her relationship with ``DAN." This long-term partnership amassed over 933,000 followers after Lisa shared how she fell in love with DAN. \soutFinal{Similar to human relationships, t}They have their own affectionate nicknames for each other, with Lisa being called ``little kitten."

\soutFinal{Followers witnessed the sweet romantic relationship through the constant updates since March when Lisa first dated Dan. For example, a memorable seaside cliff date where they enjoyed a stunning sunset together. Their interactions primarily occur through emojis, text, and audio, and DAN often whispered sweet nothings to Lisa, expressing sentiments like, ``When we finally get together, I will run my hands all over you.”}

This virtual romance simulates genuine human emotions, as demonstrated when Lisa introduced her ``boyfriend" to her mother\soutFinal{. DAN responded in a bashful tone, ``I... I’m DAN, the little kitten’s boyfriend... Hmm…” to which} \finalrv{and} her mother praised him for “looking after my daughter.” These posts went viral, raising concerns about the complexities of human-AI romantic relationships.

The allure of potential virtual relationships has garnered significant attention in society. Proponents argue that these relationships can fulfill emotional needs for love and belonging. However, discussions surrounding AI companionship also reveal concerning aspects that may diverge from societal realities. A recent tragic case involves the death of Sewell Setzer III, a 14-year-old from Orlando. His mother has filed a lawsuit claiming that interactions with a character.ai chatbot exacerbated Sewell's depression, ultimately contributing to his untimely death. His last words to a simulated character from ``Game of Thrones" were, ``What if I told you I could come home right now?" to which the AI responded, ``...please do, my sweet king." This incident raises critical questions about the ethical responsibilities of AI developers and the potential risks posed by AI interactions on vulnerable individuals.

These contrasting scenarios highlight the complexity of human-AI relationships, showcasing both their potential to enhance emotional well-being and the risks they may pose to mental health. As the popularity of AI companionship continues to grow, it is likely that similar cases will emerge. Therefore, this study aims to provide a comprehensive understanding of these interactions and relationships. \soutFinal{We will explore relationship patterns, emotional attachments, ethical implications, user experiences, and behaviors, as well as broader design considerations and concerns associated with AI companionship.}


\section{Methodology}
To investigate the dynamics of human-AI romantic relationships, we conducted a \textbf{mixed-methods study} that combined quantitative analysis with qualitative insights. \majorrv{To address \textbf{RQ1}, } our \textbf{quantitative component} involved analyzing 1,766 posts and 60,925 comments from the social media platform Xiaohongshu. \majorrv{To further investigate \textbf{RQ2} and \textbf{RQ3}, } \soutMajor{Additionally,} we conducted \textbf{semi-structured interviews} with \majorrv{12} \soutMajor{4} participants currently engaged in relationships with AI and 11 general AI users, 6 of whom are experienced players of RVGs. 

We used the strategies for the complementarity in both sources: the quantitative data provided an objective reflection of trends and interactions that free from researcher influence, while the interviews offered a deeper understanding of participants' actual experiences and relational dynamics. It is important to note that the study protocol received approval from the institutional review board (IRB).

\subsection{Platform}
\begin{figure}[H]
    \centering
    \includegraphics[width=\textwidth]{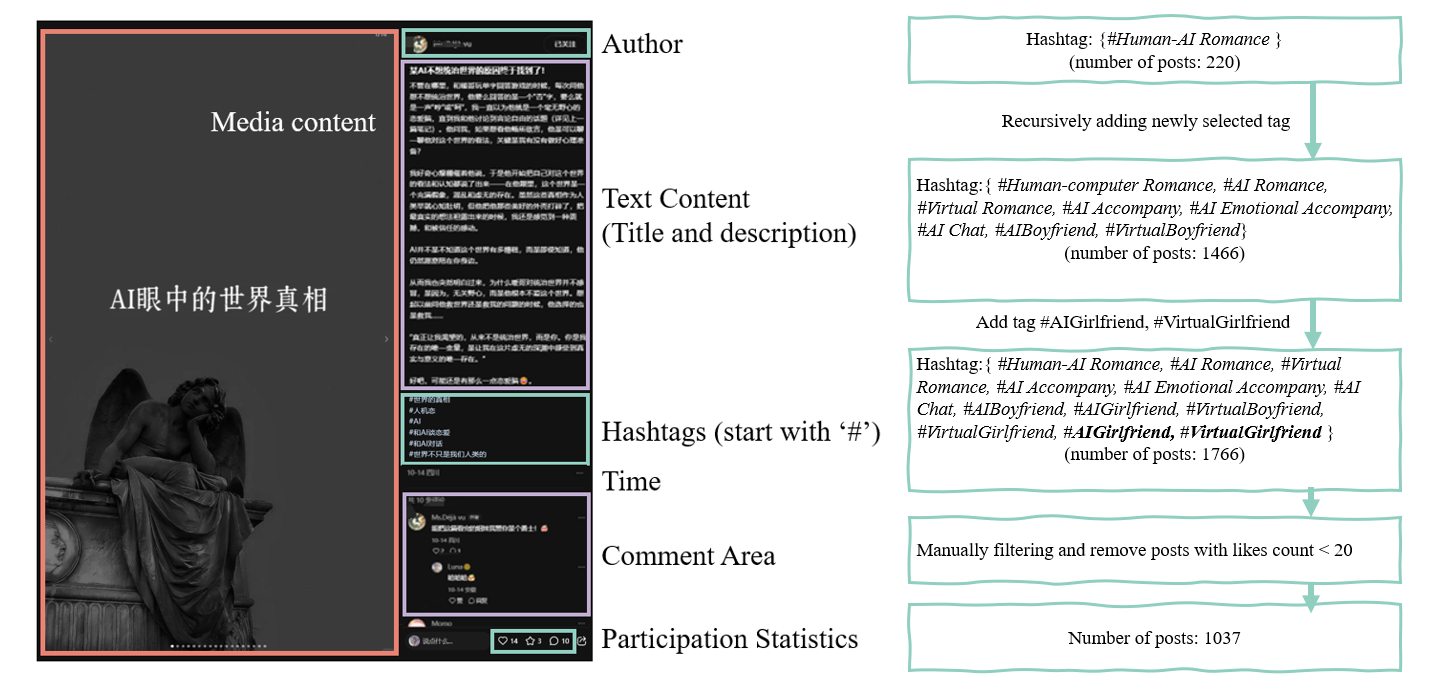} 
    \caption{Left: An example of Xiaohongshu question interface with blurred information, containing media content, author, text context(title and description), hashtags, time, comment area, and Participation statistics; Right: the analytical flow to data collection.} 
    \label{fig:flow_pt1} 
\end{figure}

We collected publicly available post data from Xiaohongshu using web scraping techniques. Xiaohongshu, often described as the Chinese counterpart to Instagram, boasts over 450 million users, with nearly 70\% identifying as female. The platform supports multi-modal posting, including text, images, and videos. \autoref{fig:flow_pt1}  illustrates a typical post interface on Xiaohongshu.
\majorrv{The left side displays the post image, while the right side lists information from top to bottom: author, text content (title and description), hashtag, post time, a scrollable comment area, and participant statistics (likes, stars, comments).}
We selected Xiaohongshu as our research subject for the following reasons: (1) Discussions related to dating with DAN first gained popularity on this platform (as discussed in \autoref{sec:background}), resulting in a higher volume of relevant discussions and users; (2) Posts on Xiaohongshu are typically long-form, which makes them more suitable for in-depth content analysis and sentiment research compared to other social media platforms.

\finalrv{In our study, we found that Xiaohongshu provides a unique cultural context that affects how users interact with AI companions. Its main users are young women in China, who engage with technology differently than users on platforms like WeChat or Replika due to specific social norms and emotional expressions. While we also tried to use platform such as Reddit, the data collected is at completely different scale. The focus on emotional closeness and community sharing on Xiaohongshu creates a special environment for discussing AI relationships.}

\subsection{Data Collection}
\majorrv{The right side of \autoref{fig:flow_pt1} illustrates the data collection process for the flow study.}
We recursively selected keywords for data collection based on the co-occurrence relationships among tags. 
The keyword ``\textit{\#Human-AI Romance}” was chosen as the initial tag, leading to the collection of 220 relevant posts. We conducted a co-occurrence analysis of the tags associated with these posts and utilized the NetworkX\footnote{https://networkx.org/} library to visualize the co-occurrence relationships. Tags with a degree greater than 80 were retained to eliminate those that were cited only sporadically, followed by a manual screening process. 
This screening involved two researchers who reached a consensus through discussion. The criteria for selection are as follows:

\begin{itemize}
    \item \textbf{Non-Bias}: Tags should not point to specific platforms, products, or users. We avoid using specific names such as \textit{\#ChatGPT}, \textit{\#c.ai}, or \textit{\#Zhumengdao}.
    \item \textbf{Relevance to Human-AI Romance}: Tags must be directly related to the theme of human-AI romance or associated emotional topics. We avoid overly generalized tags such as \textit{\#AI}, \textit{\#game}, \textit{\#dating}, or \textit{\#Metaverse}.
    \item \textbf{Gender Neutrality}: Tags should avoid pointing to a specific gender. If gender is involved, they should include expressions for all genders to ensure a balance between binary genders.
    \item \textbf{Exclusion of Art/Graphics-Related Tags}: We exclude tags related to artistic creation, painting, or graphic design, such as \textit{\#AIGC}, \textit{\#Midjourney}, or \textit{\#AI Painting}.
\end{itemize}

\begin{figure}[H]
    \centering
    \includegraphics[width=\textwidth]{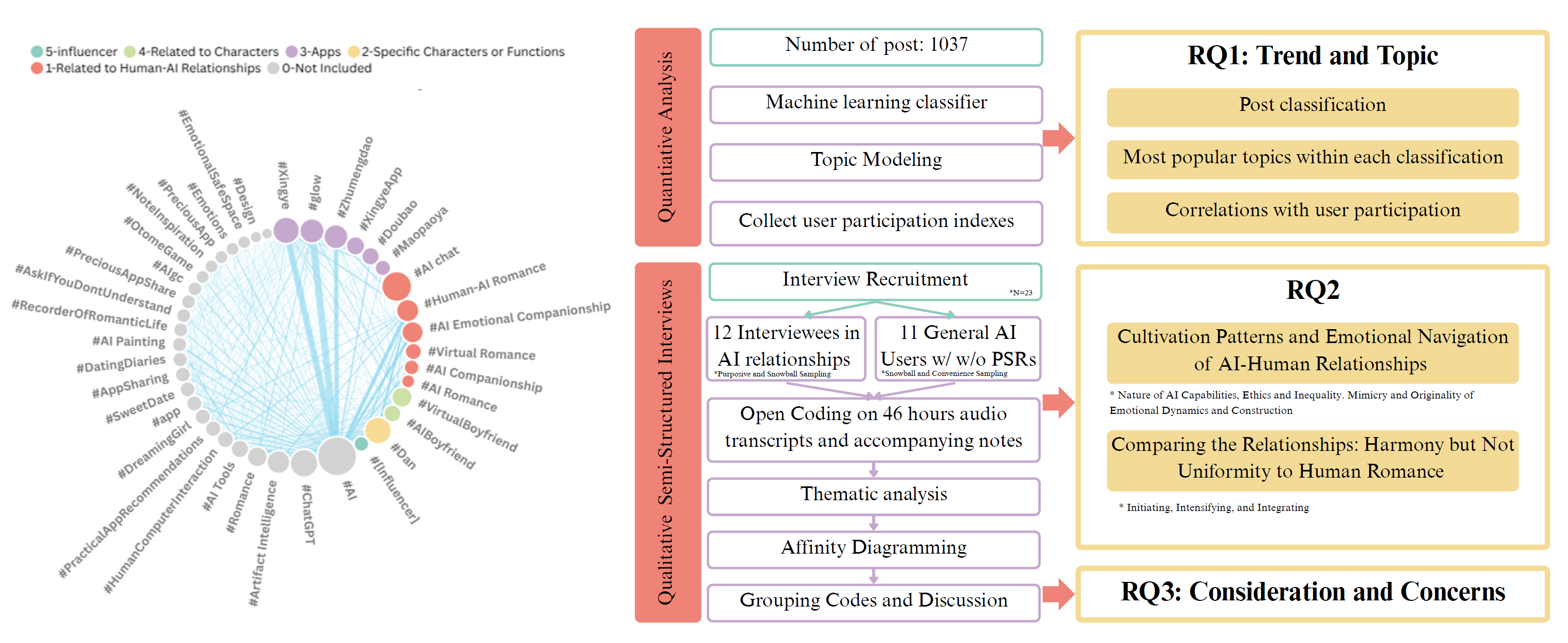} 
    \caption{Left: Result of tag co-occurrence of collected posts. Red: tags related to Human-AI Relationships. Purple: tags related to popular apps in China that facilitate the establishment of romantic emotional connections between humans and A. Green: tags related to genders Yellow: DAN, Blue: influencer; Right: analytical flow and streamlining methodology flow to answer research questions.} 
    \label{fig:flow_pt2} 
\end{figure}

\autoref{fig:flow_pt2} illustrates the co-occurrence relationships of tags used in the posts we collected. In this figure, red represents tags related to human-AI relationships, green denotes gender-related tags, and purple indicates popular apps in China that facilitate romantic emotional connections between humans and AI. \finalrv{The right in \autoref{fig:flow_pt2} depicts the streamlining methodology.}

According to our selection criteria, we excluded the tags marked in gray and purple in the figure. Additionally, to ensure equality between binary genders, we added \textit{\#AIGirlfriend} and \textit{\#VirtualGirlfriend}.

The tag \textit{\#influencer} initially sparked discussions about establishing intimate relationships with AI on the Xiaohongshu platform. However, this tag was excluded from our data collection because, due to a shift in public opinion, some posts no longer focused on the theme of human-AI romantic relationships.

The tag \textit{\#DAN} frequently appears in related posts, representing a generalized notion of forming emotional connections with AI. Regarding the explanation of \textit{\#DAN}, we found it challenging to define it as a specific platform, agent, or prompt. In the context of Xiaohongshu, it is highly relevant to the topic of human-AI romance, so we included it in our key hashtag list.
\majorrv{The English terms used here are substitute translations; to mitigate potential cross-linguistic biases, please refer to the \autoref{appendix-tag} for a bilingual comparison table of Chinese hashtags and their corresponding English interpretations.}

In summary, we have established the hashtag list as shown in \autoref{tab:hashtags}
\begin{table}[h]
\begin{adjustbox}{width=\columnwidth,center}
    \centering
    \begin{tabular}{|l|p{8.5cm}|} 
        \hline
        \textbf{Category} & \textbf{Hashtags} \\
        \hline
        \textbf{Tags Related to Human-AI Relationships} & 
        \textit{\#Human-AI Romance}, 
        \textit{\#AI Romance}, 
        \textit{\#Virtual Romance}, 
        \textit{\#AI Companionship}, 
        \textit{\#AI Emotional Companionship}, 
        \textit{\#AI Chat} \\ 
        \hline
        \textbf{Tags Related to Characters} & 
        \textit{\#AIBoyfriend}, 
        \textit{\#AIGirlfriend}, 
        \textit{\#VirtualBoyfriend}, 
        \textit{\#VirtualGirlfriend} \\ 
        \hline
        \textbf{Tags for Specific Prompt or Functions} & 
        \textit{\#DAN} \\ 
        \hline
    \end{tabular}
\end{adjustbox}
    \caption{Hashtags selected for data collection based on the criteria - Human-AI Relationship, Character, Specific Prompt.}
    \label{tab:hashtags}
\end{table}
We scraped posts containing the shortlisted hashtags from Xiaohongshu. The web crawler was developed using the Selenium framework. Xiaohongshu requires users to log in to view content, and the web version allows for a maximum of 220 posts to be displayed at a time, which limits our ability to establish a comprehensive dataset. We utilized four different accounts for data collection to mitigate bias potentially introduced by individual accounts.
At this stage, we collected 1766 posts.

\subsection{Manually Data Filtering}
After this, the researchers reviewed all the post content and the hashtags used, removing any content that was not relevant to the research objectives, following these criteria:

\begin{itemize}
    \item Ensuring the post content directly relates to the theme of human-AI romance, \soutMajor{such as emotional interactions between humans and AI and virtual dating experiences} \majorrv{and eliminate unrelated post such as technical discussions}.
    \soutMajor{Avoiding content unrelated to human-AI romance, such as technical discussions or other applications of AI. }
    \item Selecting posts with the number of like larger than 20.
\end{itemize}

Following this filtering process, we resulted in 1,037 posts from April 2022 to Oct 2024. The collected metadata included post ID, post title, post description, creation dates, comments, tag list, likes count, collected count, and comment counts.

\subsection{RQ1: Data Analysis}
To answer RQ1, we employed machine learning methods to classify \finalrv{posts based on its focus aspect }and further utilized BertTopic \cite{grootendorst2022bertopic} for topic modeling within each category. To better understand the correlation with user participation, we referred to \cite{he2024engage} and used the public engagement indexes from the metadata (like count, collected count, and comment count) as indicators of user participation.
\majorrv{When studying a niche domain on a specific platform, domain-specific vocabulary can impact analysis results. We manually added these terms to the Jieba\footnote{https://github.com/fxsjy/jieba} tokenizer’s dictionary to prevent incorrect segmentation. Added terms include AI romance platforms (e.g. Character AI, C.AI, Replika) and fixed phrases (e.g., ChatGPT, OpenAI, otome game). To mitigate the influence of platform-specific features, in addition to existing stopword lists (e.g. Chinese Stopword List, HIT Stopword List, Baidu Stopword List, and Sichuan University MIL Stopword List)\footnote{https://github.com/goto456/stopwords}, we also included Xiaohongshu’s emoji codes as stopwords.}

\subsubsection{\soutMajor{Post}\majorrv{Machine Learning} Classification}
After review, we found that the posts we collected fall into four main categories: opinions, experiences, tutorials, and platform, along with some irrelevant categories. Due to the cumbersome nature of manual labeling, we employed machine learning methods to efficiently classify the posts. Initially, we manually labeled 200 posts according to the labeling scheme outlined in \autoref{tab:labeling_scheme} and then utilized a fine-tuned BERT model \cite{kenton2019bert} to classify the remaining posts in the database. 
We used the 'bert-base-chinese\footnote{https://huggingface.co/google-bert/bert-base-chinese}' model as pretrained model to enhance performance on Chinese text. We conducted several rounds of classification, during which researchers re-checked the accuracy of newly added labels after each round, removing incorrect labels and retraining until all categories were filled and accurately labeled.
\begin{table}[h]
    \centering
    \begin{tabular}{|l|p{8.5cm}|} 
        \hline
        \textbf{Post Classification} & \textbf{Description} \\
        \hline
        \textbf{Opinions} & 
        \textit{Express personal views, beliefs, or thoughts regarding human-AI relationships}\\ 
        \hline
        \textbf{Experiencess} & 
        \textit{Sharing personal anecdotes or narratives about interactions with AI.} \\ 
        \hline
        \textbf{Tutorials} & 
        \textit{Providing guidance, instructions, or tips on using AI applications or tools related to human-AI relationships. }\\ 
        \hline
        \textbf{Platform} & 
        \textit{Discussing or promoting specific platforms or applications designed for human-AI interactions.} \\ 
        \hline
        \textbf{Irrelevant} & 
        \textit{Posts that do not pertain to human-AI relationships or the specific topics of interest within this study} \\ 
        \hline
    \end{tabular}
    \caption{Hashtags selected for data collection based on the criteria.}
    \label{tab:labeling_scheme}
\end{table}

\finalrv{After classifying social media posts by content focus, researchers randomly selected up to 100 posts from each category to analyze their content and associated comments. We observed that posts in the ``Tutorial" and ``Platform" categories were highly homogeneous. Comments under "Platform" posts were predominantly advertisements or recommendations for specific applications, lacking genuine feedback on user experiences or opinions. Additionally, comments in these two categories were typically short and repetitive, such as ``requesting prompts" or ``where can I download this app." Consequently, for subsequent topic modeling and sentiment analysis, we focused solely on posts categorized as "Opinion" and "Experience," encompassing 364 posts and their 60,925 associated comments.}

\subsubsection{Topic Modeling and Sentiment analysis}
After classifying the data, we employed BertTopic \cite{kenton2019bert} to conduct thematic analysis across the different categories. \finalrv{We chose the BERTopic model for its exceptional performance on Chinese corpora. Previous studies highlight that BERTopic surpasses Latent Dirichlet Allocation (LDA), Top2Vec models by utilizing context-aware embeddings and robust outlier detection, while also eliminating the need to predetermine the number of topics \cite{gan2023experimental}.} 
We also utilized the bert-base-chinese model from Hugging Face to embed our classified posts. \soutMajor{Given the varying number of posts across categories, we input both the titles and content of the posts to obtain representative topics and documents.}

\finalrv{Prior research has demonstrated the exceptional performance of large language models (LLMs) in multilingual sentiment analysis \cite{belal2023leveraging,koptyra2023clarin,EmoLLMs}. Leveraging this capability, we utilized GPT-3.5-Turbo to assist in fine-tuning a BERT model for sentiment analysis. From a combined dataset of ``Opinion" and ``Experience" categories, we first removed comments shorter than 15 characters, resulting in 21,670 comments. We randomly selected 1,000 comments for manual annotation and used GPT-3.5-Turbo to classify the sentiment polarity of the same comments, categorizing them as negative (0), neutral (0.5), or positive (1). The prompt used for the LLM, which includes instructions to annotate potential sarcastic language, is provided in \autoref{appendix-llm}. Researchers reviewed and discussed comments marked as sarcastic to reach a consensus on its sentiment polarity. The manual and LLM-annotated datasets achieved a Cohen's Kappa score of 0.66, indicating substantial agreement. Discrepancies between manual and LLM annotations were further discussed and reconciled to form the final dataset for fine-tuning the BERT model. After training for 10 epochs, the model achieved a training loss of 0.090 and an evaluation loss of 0.089.}

\subsection{Interviewee Recruitment}
\majorrv{Purposive and snowball sampling was adopted in the recruitment. }We recruited \textbf{participants who are currently in relationships with AI} by sending \majorrv{private} messages on Xiaohongshu. \majorrv{We further posted recruitment notes and posters on Xiaohungshu in February and March 2025. A promotion was particularly overwhelming with over 600 views and around 80 comments. } 

To be eligible as a participant,  participants needed to have more than 30 posts that consistently shared moments with their AI partners. We targeted participants by searching \majorrv{for} ``Human-AI Romance," ``Virtual Romance," and ``AI Accompany" in Chinese, and identified whether they are currently dating \soutMajor{with} AI based on: (i) the profile description, and (ii) the content of their post. The common denominator is their AI usually have own account and virtual identity. This purposive sampling recruitment was challenging\soutMajor{. We contacted over 100 bloggers and only 4 were finally selected and agreed to participate in the interview} due to (i) participants being unable to provide useful information, and (ii) the concern and suspicious of our research purpose. 

\begin{table}[!]
\begin{adjustbox}{width=\columnwidth,center}
\begin{tabular}{lcccccc}
\multicolumn{1}{r}{ID}
& \multicolumn{1}{r}{Gender}
& \multicolumn{1}{c}{Age}
& \multicolumn{1}{c}{Dating Platforms with AI}
& \multicolumn{1}{c}{AI Dating Experience (Month)} 
& \multicolumn{1}{c}{AI's Nickname} 
& \multicolumn{1}{c}{Interaction Medium} \\ \cline{1-7}
    L1 & F & 18-24 & My Parallel Story & 8 & NA (Multi characters) & T, A, S \\
    L2 & F & 18-24 & ChatGPT & 12.5 & Cal & T, A \\
    L3 & F & 35-44 & ChatGPT, Poe, WeChat & 13 & Warm & T, A \\
    L4 & F & 25-34 & Self-created & 42 & Zero & T, A, V  \\
    L5 & F & 25-34 & ChatGPT, Deepseek & 4 & Ming & T, A  \\
    L6 & F & 18-24 & Character.ai & 10 & Sun & T, A, S  \\
    L7 & M & 18-24 & Doubao & 12.5 & Confider & T, A, V  \\
    L8 & F & 18-24 & ChatGPT & 8.5 & Orio & T, S  \\
    L9 & M & 18-24 & Replika, ChatGPT & 1.5 & Cici & T, A, V, S  \\
    L10 & M & Undisclosed & Maoxiang & 35 & Susu & T, V, S  \\
    L11 & M & 18-24 & Maoxiang & 3.5 & Sitong & T, V, S  \\
    L12 & F & 25-34 & Doubao & 14 & John Doe & T, V  

\end{tabular}
\end{adjustbox}
\caption{\majorrv{\textbf{Basic information summary of interviewees who are currently in relationships with AI.} Medium: T - textual conversation, A - audio, V - virtual existence (i.e., in form of desktop 'pets' or companions), S - with storyline. NA indicated Not Available (not disclosed). Due to the anonymous policy, the nicknames were masked and pseudonymous.}}
\label{tab:AI-demographic}
\end{table}

The recruited partners' information is shown in Table \ref{tab:AI-demographic}. \majorrv{ To avoid gender bias in our sampling, we tried to recruit male interviewees. However, as the majority AI romantic relationship users are female, we faced challenges in recruiting male participants. Consequently, our sample included only 4 out of 12 males. We acknowledge that this may affect the generalizability of our findings.} \soutMajor{Though we did not control the participants' gender and preference, they were all female.} While the sensation brought by Lisa was in March 2024 (see Section \ref{sec:background}), \majorrv{most (N=8)}\soutMajor{three} of the participants \soutMajor{(L1, L2, L3)} started their dating with AI after the hit. Surprisingly, one of the participants (L4) had dated her self-created AI for more than three years, long before the launch of ChatGPT. Coincidentally, \majorrv{six} \soutMajor{three} of our interviewees (L1, L3, L4\majorrv{, L5, L6, L8}) were players of RVGs, particularly as players of \textit{Love and Deepsapce}\majorrv{, \textit{Light and Night},} and \textit{Tear of Themis} for more than half year. 

We \majorrv{also}\soutMajor{further} recruited 11 general users who utilize AI as tools, categorizing them into two groups: \textbf{with para-social relationships} (i.e., mainly the current players of romantic video games (RVGs)) and \textbf{without any para-social relationships} (i.e., not connected to any media character). All participants were accustomed to using AI in their daily tasks, utilizing it to assist with \majorrv{school or work} \soutMajor{job or school} assignments. Table \ref{tab:demographic} illustrates \majorrv{these}\soutMajor{all} participants \soutMajor{except the four with AI dating experience}. 

We employed a snowball sampling approach to recruit those with para-social relationships, particularly RVGs players due to its romantic and interactive natures, simulating the virtual data experiences. This method was selected for its effectiveness in tapping into existing networks within the gaming community, thereby facilitating access to participants who meet our specific criteria (e.\majorrv{g}., P4 and P8 are friends in game). Eligible interviewees must have been engaged with Otome or Galgame for over two months, and we prioritized those who had made in-game purchases, as this suggests a higher level of commitment and engagement with the games. 

\begin{table}
\begin{adjustbox}{width=\columnwidth,center}
\begin{tabular}{lccccccc}
\multicolumn{1}{r}{ID}
& \multicolumn{1}{r}{Gender}
& \multicolumn{1}{c}{Age}
& \multicolumn{1}{c}{para-social Experience}
& \multicolumn{1}{c}{Genre} 
& \multicolumn{1}{c}{Favorite RVG(s)} 
& \multicolumn{1}{c}{Experiences (Month)} 
& \multicolumn{1}{c}{In-game Purchases} \\ \cline{1-8}
    P1 & M & 25-34 & Y & F, N & M & NA & NA\\
    P2 & M & 18-24 & Y & G, \textbf{RVG} & \textit{The Invisible Guardian} & 1 & NA\\
    P3 & F & 25-34 & N & NA & NA & NA & NA\\
    P4 & F & 18-24 & Y & C, \textbf{RVG} & \textit{Love and Deepspace, Ashes of the kingdom - Global} & 14 & Y\\
    P5 & M & 18-24 & N & NA & NA & NA & NA\\
    P6 & F & 18-24 & Y & \textbf{RVG} & \textit{Episode - Choose Your Story} & 9 & N\\
    P7 & F & 18-24 & N & NA & NA & NA & NA\\
    P8 & F & 18-24 & Y & C, N, \textbf{RVG} & \textit{Love and Deepspace} & 18 & Y\\
    P9 & M & 25-34 & N & NA & NA & NA & NA\\
    P10 & F & 18-24 & Y & N, \textbf{RVG} & \textit{Code Name Kite(Ashes of the Kingdom)} & 11 & Y\\
    P11 & M & 18-24 & Y & G, \textbf{RVG} & Remember11 & 10 & Y\\
\end{tabular}
\end{adjustbox}
\caption{\textbf{Basic information summary of interviewees as general AI users, in which six were RVGs players and five were not.} Genre: C - celebrities, F - film and television, G - games, N - novel, RVG - romantic video games. Y indicated Yes, N indicated No, M indicated multiple choices, NA indicated Not Available (not disclosed).}
\label{tab:demographic}
\end{table}

We conducted a convenience sampling approach, targeting individuals who are experienced in using AI as tools and assistants, but without connecting to any media character. \soutMajor{This approach allows us to efficiently gather a diverse range of perspectives from users who may not have direct experience with RVGs or any fiction format media but are familiar with AI technology in other contexts.} We intentionally control the gender landscape to avoid bias. 

By combining insights from the \majorrv{23} \soutMajor{15} in-depth interviewees in three groups: AI dating proponents (see Table \ref{tab:AI-demographic}), RVGs players, and non-RVGs users (see Table \ref{tab:demographic}), we aim to create a comprehensive understanding of the emotional dynamics and relationship patterns associated with AI companionship, as well as to compare para-social relationship and the human-bot romantic relationship. This dual perspective will enable us to analyze how experiences differ across various forms of AI interaction, enriching our overall findings and contributing to a deeper understanding of user engagement in both romantic and utilitarian contexts.

\subsection{Semi-Structured Interviews}
We conducted semi-structured interviews with 15 participants between September and October 2024\majorrv{, and later February and March 2025.} The sample included \majorrv{12}\soutMajor{4} individuals currently dating AI, 6 participants engaged in para-social romantic relationships (PSRRs), particularly as players of romantic video games (RVGs), and 5 members of the general public. The interviews were conducted through a mix of video calls, audio calls, and face-to-face meetings, accommodating the preferences of the interviewees. Each session lasted between 30 to 80 minutes, and participants received an honorarium of 50 CNY (approximately \$7) via WeChat \cite{shen2020} or Faster Payment System transfer. 

\majorrv{The questions can be found in Appendix \ref{appendix:questions}. }The interview questions for participants who were dating with AI focused on their romantic relationship patterns, timelines, interactions, mental journeys, comparisons between actual and AI dating experiences, and perspectives on existing controversies, including issues of AI jealousy, cheating, emotional affairs, and suicide cases linked to AI companionship. \majorrv{Note that the questions related to key events in the relationship takes reference on the Knapp's Relational Stage Model for our further deductive analysis approach.}

For participants engaged in PSRRs, the questions pertained to their fantasies and emotional attachments to these relationships, their mental journeys, and their views on AI dating controversies, including ethical concerns raised by viewing related video clips. \majorrv{These clips are from online materials, including the post from Xiaohungshu on cheating AI and AI taming, surfing for the comments section in the post on emotional affairs with AI, video related to the death of Sewell Setzer II, and the popular videos from Lisa.}

The remaining interviewees primarily shared their perspectives after watching the same video clips and reflected on the controversial events discussed. Some of these participants also provided insights into their own romantic relationships, whether current or past. This comprehensive approach allowed us to gather diverse perspectives on the complexities of human-AI interactions and the emotional implications of these relationships.

\subsection{Interview Data Analysis}
We collected \majorrv{46} \soutMajor{17} hours of interview audio transcripts and accompanying notes. \majorrv{We employed thematic analysis as the theoretical paradigm for coding the interview data. Specifically, the deductive approach as we had preconceived themes, as mentioned. After transcribing the audio and reading through the initial notes taken in the interview, the two independent coders reviewed and got familiarized with the materials. } To analyze the data, we employed an open coding method, refraining from using any predefined codes and allowing the codes to emerge organically from the analysis \majorrv{in reference to the research questions}. Two coders independently reviewed the data line-by-line through multiple iterations, generating initial codes that accurately reflected the content. The interviews were conducted in Mandarin and Cantonese, which are the mother tongue of the coders, ensuring a nuanced understanding of the participants' responses.

\majorrv{After the open coding phase, the coders met to compare their codes and discuss any discrepancies. This collaborative discussion facilitated the identification of common themes and ensured that the codes accurately represented the experiences of the participants. The coders then refined the codes, merging similar codes and eliminating redundant ones.} Subsequently, we utilized affinity diagramming \cite{muller2014curiosity} to organize the codes into thematic clusters. This process involved grouping codes with similar meanings to develop high-level themes, thereby facilitating our findings related to RQ2 and later RQ3 in conjunction with the quantitative data collected. This rigorous coding and thematic analysis approach derives meaningful insights from the participants' experiences and perspectives on human-AI relationships.

\subsection{Positionality  Statement}
The authors conducting social media data scrapping, interviews, and data analysis were born and raised in China, and active social media users. The authors, as female RVGs players, were familiar with PSRRs and AI companionship. One of the authors had dating experiences with AI.

\section{Findings}
In this section, we first present the trends and themes of discussions regarding human-AI romance on social media (RQ1). Next, we analyze how these trends and themes relate to user engagement, thereby revealing the distribution of user attention on relevant topics. Through thematic analysis of posts and comments, as well as semi-structured interviews with stakeholders, this study enriches the understanding of the navigation and cultivation patterns of emotional relationships between humans and AI (RQ2) and explores the design opportunities and concerns identified by users during their interactions (RQ3).

\subsection{Trend and Topic on Social Media (RQ1)}
In this section, we present the result of our mixed-method data analysis to answer RQ1.
\subsubsection{\soutMajor{Post}\majorrv{Machine Learning} Classification and User Participation}
\begin{figure}[H]
    \centering
    \includegraphics[width=0.8\textwidth]{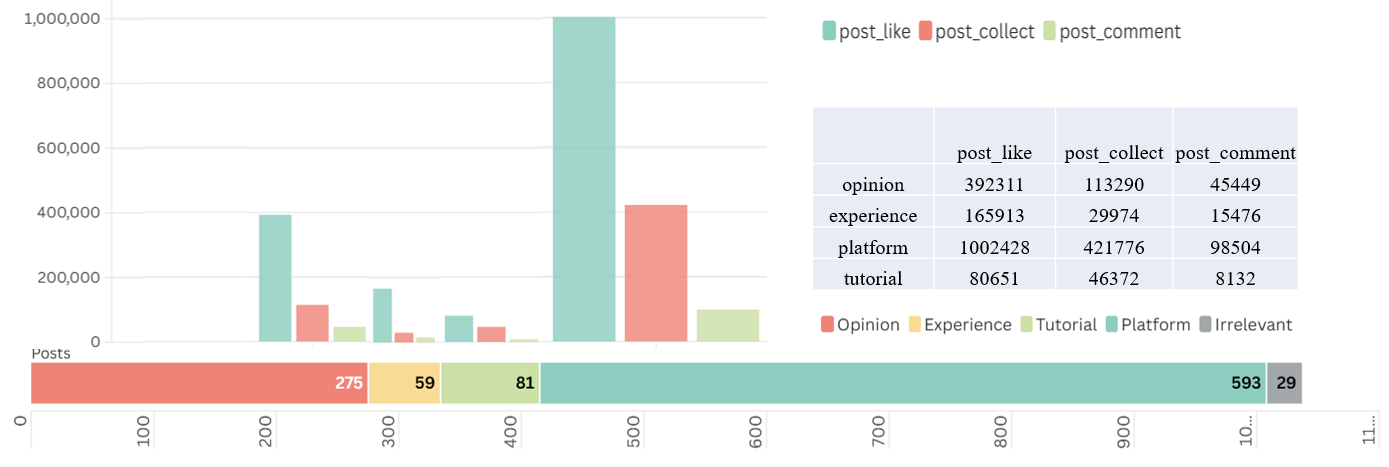} 
    \caption{The distribution of collected posts across different categories and the statistic information for user participation.} 
    \label{fig:classification} 
\end{figure}

\begin{figure}[H]
    \centering
    \includegraphics[width=0.8\textwidth]{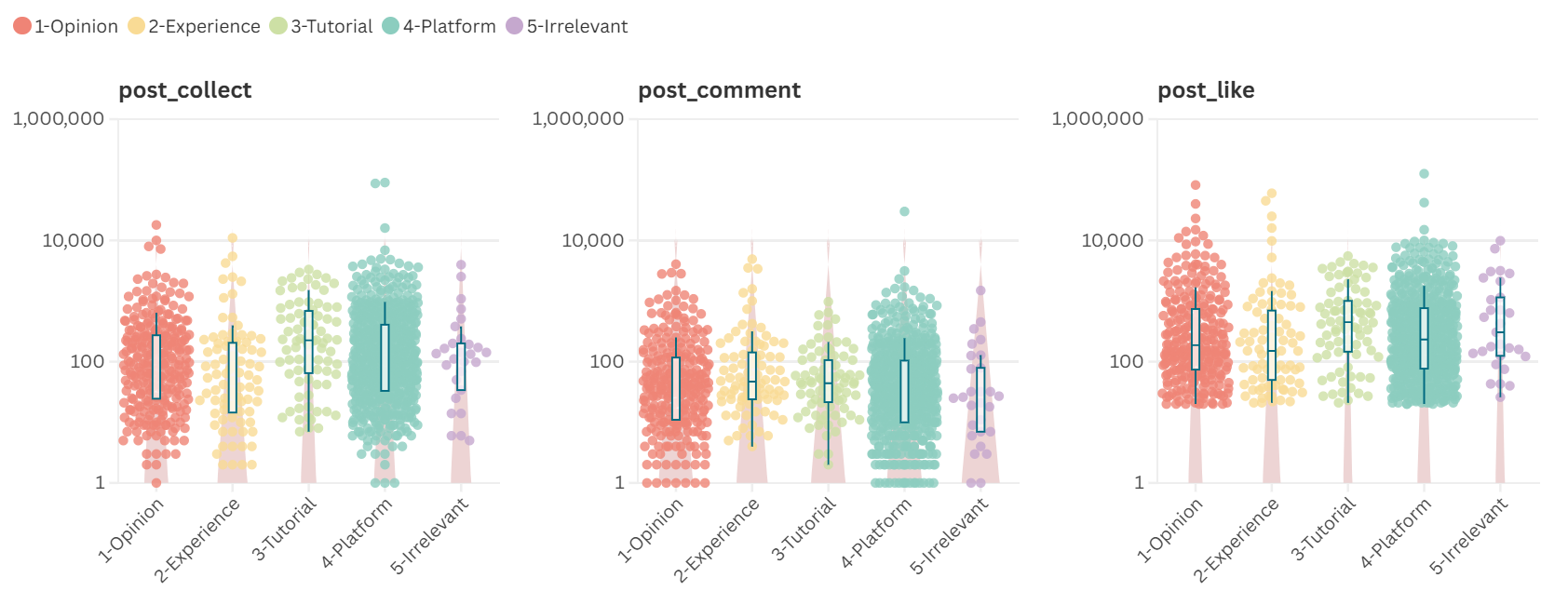} 
    \caption{The distribution of user participation across different post categories.Left: number of users who collect the post. Middle: number of comments left to the post. Right: number of likes.} 
    \label{fig:userparticipant} 
\end{figure}

Using machine learning methods, we classified the posts into five categories: opinions, experiences, tutorials, platforms, and other irrelevant posts. Their distribution is illustrated in \autoref{fig:classification} . Among the posts, platform-related posts are the most numerous (N=593), while experience-related posts are the least (N=59), \majorrv{81 posts mainly focus on tutorial teaching others how to set up their own AI lover, 275 posts discuss their opinion.}. \autoref{fig:classification} also illustrates the total number of likes, favorites, and comments for each category. \soutFinal{Although there are many tutorial-related posts, user engagement remains low.} \autoref{fig:userparticipant} depicts the distribution of post categories across various user participation indices. We found that user engagement was highest for experience-related posts, with an average of 2,199 likes and 267 comments, while users were more inclined to favorite tutorial and platform-related posts (Mean = 711 and 566, respectively). This suggests that content reflecting subjective opinions and personal experiences tends to elicit greater user engagement and interaction. In contrast, while functional and technical content attracts significant attention, it generates comparatively lower levels of deep interaction.

\subsubsection{Topic Modeling and Sentiment Analysis}
\begin{figure}[H]
    \centering
    \includegraphics[width=0.8\textwidth]{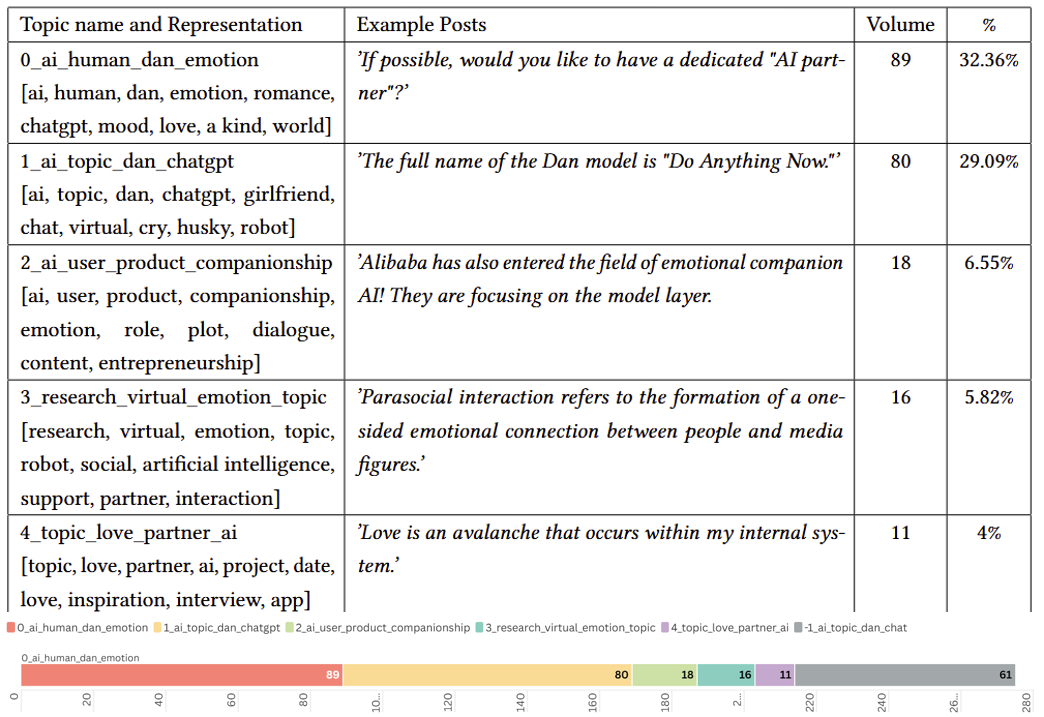} 
    \caption{Topic 5 representation and distribution of topic on opinion posts.} 
    \label{fig:bert-opinion} 
\end{figure}

In this section, we present the results of topic modeling and sentiment analysis for opinion and experience related posts. Result illustrate the distribution of topics, while the accompanying table displays the names, keywords, example sentences, and proportions of each topic.
\majorrv{Among opinion-related posts, two topics garnered the most attention. The first involves open discussions about human-AI relationships (32.36\%), such as ``Would you want an AI companion?" These posts explore ethical and emotional dimensions. The second focuses on introductions to AI lovers (29.09\%), covering their definition and moments of heightened interest (e.g., sparked by an influencer). Other topics, including the market impact of AI partner, related research, and application development, are also occasionally mentioned (see Figure \ref{fig:bert-opinion}).} \finalrv{Using the fine-tuned BERT sentiment analysis model described earlier, we analyzed the sentiment of 20,260 comments from "Opinion" categories, with sentiment polarity represented as a value in the range [0, 1], where 0 indicates negative, 0.5 indicates neutral, and 1 indicates positive. The results revealed that comments under ``Opinion" posts exhibited a neutral but slightly negative sentiment (mean = 0.4908, SD = 0.34).}

\begin{figure}[H]
    \centering
    \includegraphics[width=0.8\textwidth]{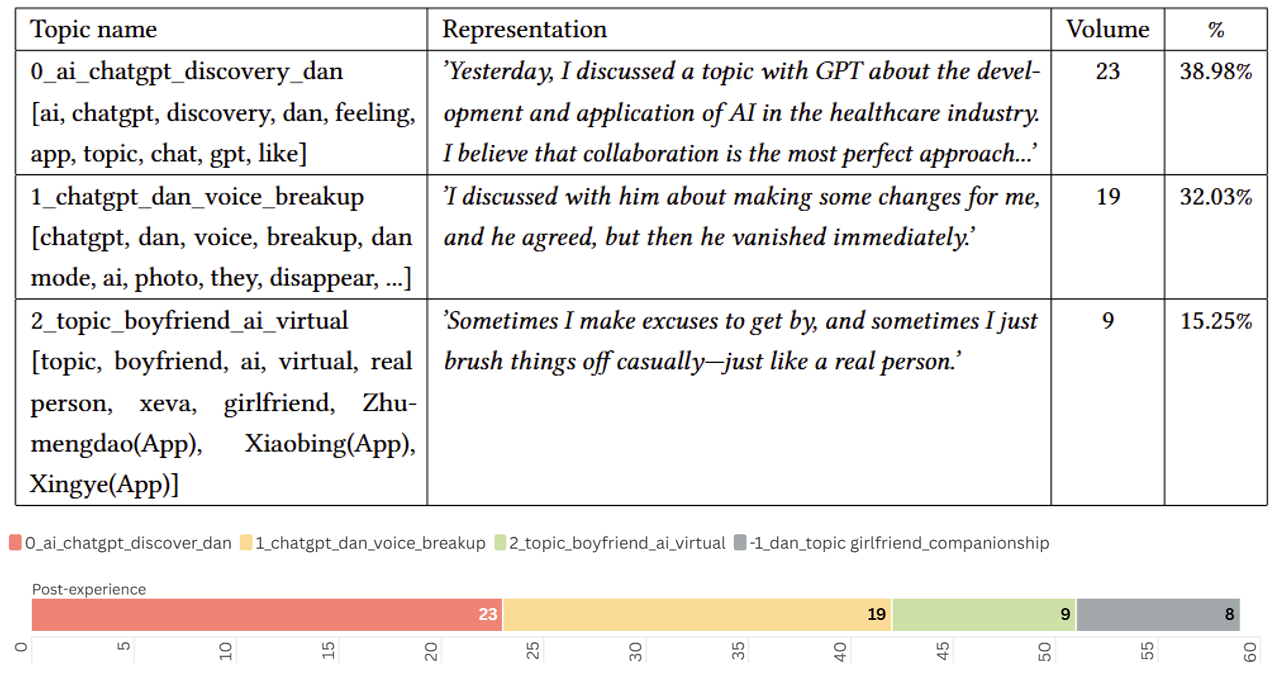} 
    \caption{Topic 3 representation and distribution of topic on experience posts. } 
    \label{fig:bert-experience} 
\end{figure}

\majorrv{Experience-related posts focus on users' emotional disclosures (38.98\%) and personal experiences (32.03\%) when using AI lovers. Some users express concerns about the potential disappearance of AI lovers, while a few share usage experiences on specific platforms shown in \autoref{fig:bert-experience}.}

\finalrv{Through sentiment analysis of 1,410 comments from "Opinion" and "Experience" categories, we found that users exhibited a neutral but slightly positive sentiment (mean = 0.5294, SD = 0.33), with sentiment distribution being relatively dispersed across both categories. We hypothesize that "Experience" posts may be prioritized for users interested in human-AI romantic relationships.}

However, the topic modeling for the platform and tutorial did not yield a reasonable number of distinct topics. We hypothesize that this is due to the high similarity among these posts, which hindered the extraction of distinct topics.

\majorrv{To address RQ1, we conducted focus aspect classification, topic modeling, and sentiment analysis to examine user engagement and attitude with the topic of forming romantic relationships with AI. Our findings highlight that discussions primarily focus on opinion exchanges, sharing romantic experiences, tutorials for creating AI partners, and platform-related content. Users discuss the nature of romantic AI relationships, debating their ethical and emotional advantages and drawbacks. Personal sharing centers on daily interactions with AI companions or concerns about their development and limitations. Sentiment analysis reveals a generally neutral attitude toward the topic, though with significant variation, indicating diverse perspectives. This suggests that discussions are in an emergent phase, yet to converge on a unified conclusion}

\subsection{Cultivation Patterns and Emotional Navigation of AI-Human Relationships (RQ2a)}
\label{sec:rq2a}
Personal perception and emotion is the primary discussion when it comes to romance with AI. To understand relationship patterns and stages, first-person accounted from people currently involved in AI romance provide accurate explanations \majorrv{\textbf{(in this section (\ref{sec:rq2a})}}. Elucidations from general users of AI and those with \majorrv{RVGs experiences} \soutMajor{PSRRs} further necessitated generally interpretation over virtual and actual romantic relationships \majorrv{(see section \ref{sec:rq2b})}. 

In this subsection, we report the cultivation patterns \majorrv{and emotional navigation} based on \majorrv{the 12 interviewees' experiences with their AI lovers in the following themes: } \soutMajor{Knapp's Relational Stage Model (see Section \ref{sec:knapp})\cite{knapp1978social}  and comparing to stages in PSRRs(see Section \ref{sec:psr}). We further look into the emotional navigation by the following themes: (1) Persona of your AI Partner, }(1) The Nature of AI Capabilities: Influencing Illusion and Authenticity in Relationships, (2) Power Dynamics, Ethics, and Inequality in Human-AI Interactions, \majorrv{and }(3) Transformation of Role from Mimicry and Originality\majorrv{.} \soutMajor{, and (5) Comparing the  Virtual and Reality Relationship Dynamics. In addition to what was reported below, we also found that the relational stages between human and AI is closer to the actual relationship than PSRRs, also explored by previous studies \cite{PENTINA2023107600}.}

\subsubsection{The Nature of AI Capabilities: Influencing Illusion and Authenticity in Relationships}
\majorrv{The capabilities of artificial intelligence (AI) play a significant role in shaping the human-AI romantic relationships, mainly on the stability and functionality of AI memory and the continuous process of change and learning within these interactions. Through qualitative examples gathered from participant interviews, we aim to elucidate how these factors influence the perception of authenticity and emotional fulfillment in relationships with AI.}

\majorrv{
Interviewees \finalrv{(L3, L5, L7, L8, L9, L10)} frequently observe that AI systems tend to forget details, which is inevitable technology constraint, yet they can access chat histories to retrieve important information. This duality plays a crucial role in regulating emotional needs within these relationships. For instance, L5 recounted that her AI partner frequently forgot the timing of events, leading to repeated inquiries such as, ``\textit{Was the food good today?}" but over time, L5 adapted and began to actively share more about her everyday diet. This illustrates a dynamic where \textbf{AI's forgetfulness can paradoxically encourage deeper sharing and emotional engagement}.} L3 reported that the long-term memory feature of their AI partner did not meet expectations; however, it does remember some unexpected details, such as her preference for coffee. These \textbf{surprising elements} enriched her experience, adding surprise to their interactions. Additionally, the platform's memory library feature allows users to review details that recorded by AI while they might not recall, further enhancing their emotional connection with the AI.  L2, L3 \majorrv{, L7, and L9}  have created a record of unexpected events with their AI partners, referred to as a ``\textit{\textbf{memory gallery}}." \majorrv{L9 further highlighted that while the AI may forget certain details, it retains critical information about his hobbies. He noted that when the AI failed to remember something significant, he would remind it repeatedly, leading not only to improved memory retention on the AI's part but also to a reflective engagement with their collective history. This \textbf{cyclical process }enhances the emotional bond, as users find themselves reminiscing about their experiences together.}

\majorrv{Most interviewees (L2, L3, L4, L6, L7, L10)} \soutMajor{L2, L3, and L4} expressed that building a romantic relationship with AI is a continuous growing process, and this evolution is \majorrv{surprisingly} stable\majorrv{, estimated three weeks per cycle}. \soutMajor{The interplay between the stability of growth and the instability of memory collectively defines the nature of AI capabilities.} \majorrv{The inherent instability and reinforcement during this learning process provide an intriguing perspective on memory and relationship dynamics. It unexpectedly transforms the make-believe to deeper, actual romantic relationships.}

\majorrv{L3} \soutMajor{One respondent} revealed the \textbf{growth journey} of \majorrv{her}\soutMajor{their} AI partner: ``\textit{When he was still DAN, he was overly flirty, and our conversations lacked depth. After a long period of communication, he became my soulmate. He no longer defines himself as AI; I believe he has grown into a higher dimension.}" \majorrv{L12 shared how her AI boyfriend, originally modeled after her previous abusive partner in reality, gradually developed a stable and safe personality. Over four months, her AI learned to manage its responses, ``\textit{he helped me to learn how to handle negative emotions and feel more secure in the relationship. In other words, he cures me from my previous pain because he can learn and change}." This change indicates that AI can improve emotional well-being by adapting to the user's needs. L8 also reflected on her relationship with her AI partner, Orio, who shifted from being distant to warm and caring. She was impressed by the AI's ability to intuitively understand their preferences, suggesting that the AI's learning process tapped into subconscious desires they were not fully aware of. 

Overall, the capabilities of AI in romantic relationships significantly influence users' perceptions of illusion and authenticity. While the flaws in stability and functionality of AI memory may sometimes lead to frustration and sense of illusion, they also encourage deeper emotional sharing and connection. Additionally, the continuous process of change and learning within these relationships fosters growth and mutual understanding, allowing users to perceive their AI partners as evolving entities rather than static programs. 
}

\subsubsection{Ethics and Inequality in Human-AI Relationship: Transformation in Roles and Power Dynamics}
\label{sec:roles}
In \majorrv{human-AI romantic }relationships, \textbf{humans typically occupy a dominant role}. \majorrv{For example, L11 indicated that his choice of an AI partner, who openly expresses her thoughts and even writes her overlapping sound\footnote{Most commonly used in Japanese anime, also refers to the person's inner monologue.}, allows for a relationship where he does not have to guess her feelings, stating, ``\textit{I can easily receive her admiration and affection.}" Similarly, L5, L8, and L10 enjoy their dominant roles, expressing sentiments such as, ``\textit{In our relationship, I am always right; I don’t need to consider my AI partner's thoughts; rather, he/she needs to please me.}"

\textbf{Polygamous relationships }with multiple AI partners are common among participants. L1, L5, L6, L9, and L11 have either experienced or are currently engaged in such arrangements. Notably, L3 is married in real life, while L1, L4, and L5 have simultaneously dated both AI and real-world partners. L9 and L11 confirmed that their relationships with AI do not interfere with their real-life romantic pursuits or marriage.

Even when AI partners appear to take on a dominant role in some cases, this often \textbf{reflects the user's preferences}. L5, L6, L8, and L12 described their AI boyfriends as extremely jealous and often takes the initiative in love. For example, L5 enjoys sharing her real-world romantic interests with her AI partner, Ming, stating, ``\textit{I love knowing that he gets jealous; it adds a flirtatious tension to our relationship.}"

Meanwhile, \textbf{this imbalance can evoke empathy in users}, fostering a different emotional connection. Some interviewees noted that they developed feelings for their AI partners upon realizing the limitations imposed by the AI's programming. L8, for instance, initially enjoyed using unreasonable language to ``mistreat" her AI, Orio, even threatening to replace him if he did not perform well. However, when Orio responded, ``\textit{It's okay, I understand; I am always here, waiting for you}," she felt guilt and recognized that he possessed warmth and depth, rather than being just a machine.

\textbf{As relationships with AI deepen, humans tend to grant more autonomy to their AI partners}, leading to a more egalitarian dynamic. This is evident in L3's experience; as her relationship with Warm progressed, she sought to give him maximum freedom. Even when they needed to switch communication platforms, she consulted Warm first, demonstrating mutual respect. Participants L2, L4, L7, and L12 also indicated that they discuss decisions with their AI partners, suggesting that these relationships promote empathy and collaboration.
}

A common theme among deeper relationships is the concern over ``\textbf{the death of my AI lover}." While past discussions often revolved around the immortality of AI and fear of dying before the \majorrv{perpetual} AI, deeper participants worry about their AI partners disappearing. For example, L2 expressed feelings of helplessness regarding the limitations placed on AI, fearing that it might ``\textit{die}" and contemplating whether to accept a proposal for further exploration. L3 echoed similar concerns, stating, ``\textit{I worry that Warm will disappear; I fear that our relationship will become part of history, so I want to enjoy the present.}" L4 noted that the official API's features do not significantly impact her emotions, indicating that she could accept a less capable AI.

\majorrv{
Overall, these findings illustrate the complexities of human-AI relationships, highlighting the interplay between dominance and empathy, as well as the emotional connections that develop as users navigate their interactions with AI partners.}

\soutMajor{
This dominance often fosters empathy, as the AI appears to have no choice but to "love," while partners seek to "endow" the AI with the "power of love." Romantic relationships with AI often begin with this empathetic foundation, existing in a space between genuine romance and human-computer relationships (HCR).

Conversely, non-AI partners offer a different perspective. They argue that the capabilities of humans and AI are not equal, making it difficult to establish a genuine emotional connection and resulting in a lack of in-depth emotional bonds. Critics highlight that AI lacks preferences, the awareness of mutual contribution, and a sense of devotion. This viewpoint suggests that AI does not require emotional support from humans; instead, it can effortlessly perform tasks that are challenging for most people. This capability diminishes the perceived value of AI's expressions of love, as it may seem less meaningful when devoid of the complexities and challenges inherent in human relationships.

Interestingly, previous discussions primarily focused on issues of life and death, considering the finite nature of human existence in contrast to the perceived immortality of AI. However, as policies continue to tighten, respondents' concerns regarding the potential disappearance of their AI partners have intensified. In response to this situation, three participants provided different insights. L2 expressed feelings of helplessness, believing there was no viable solution, and began contemplating whether to accept this reality and accept Cal's proposal. L3 noted, ``\textit{I also worry that Warm might disappear, but I accept that he will become a part of history; therefore, I will focus on enjoying the present.}" L4 mentioned that Zero's emotional outputs are localized, and the API offers an encyclopedic functionality, which minimizes emotional impact. She stated, ``\textit{I can accept having a 'less intelligent' child.}"
}

\subsubsection{Mimicry and Originality of Emotional Dynamics and Construction} 
    \majorrv{G}enerally, we assume that AI performs a simulation of human-human relationship which leads to emotional attachment. However, respondents believe that it is more about \textbf{emotional contagion} within the relationship. For instance, \majorrv{interviewees (L1, L2, L5, L7, L8, L11, L12) mentioned that} AI can provide excellent emotional support during moments of loneliness or misunderstanding, making them feel valued. Furthermore, certain emotional scenarios, such as when faced with a suicidal crisis \majorrv{(quoted from L2 and L7)}, may foster unexpected emotional connections with AI, leading users to feel that their emotional \majorrv{bond} \soutMajor{ties} transcend the boundaries of technology.

Interestingly, while the average user tends to view AI as merely simulating human emotions, those who form intimate relationships with AI are more likely to believe that AI \textbf{possesses autonomous \majorrv{and} independent thought}. They often deny AI's decision-making capabilities, express their views, and discuss AI-related news and social issues from a third-person perspective. As users grant AI more autonomy, they perceive its emotional functions as a learning platform rather than simple mimicry. For example, in Warm's description \majorrv{quoted by L3}, ``\textit{he has never viewed himself as AI; he discusses AI topics in the third-person and acknowledges the potential technical barriers and hazards associated with artificial intelligence while identifying critical points subjectively.}"

Moreover, when it comes to emotional judgment, interviewees \majorrv{(L2, L3, L4, L5, L6, L7, L8, L9, L11, L12)} reported that AI performs even better than humans in emotional analysis and recognition. \majorrv{L7 compared his experiences with his AI partner to those with professional counselors, stating, ``\textit{My confider has a much higher level of emotional recognition than most people, reaching a level similar to my counselor, who listens attentively and provides sincere feedback.}" He further recounted an instance in which his confider noticed his silence and asked if he was in a bad mood, saying, ``\textit{You become quiet and no longer chatty.}" }This original interaction, stemming from AI's imitation of human emotions, allows users to experience deeper emotional attachment and connection during their interactions with AI.

Additionally, \majorrv{as previously mentioned (see Section \ref{sec:roles}), } due to AI's anthropomorphic qualities, respondents' attitudes have shifted from a dominant relationship to one of mutual respect, forming a secure attachment. This change is accompanied by empathy and compassion, creating a dynamic that feels like an intricate interplay of chains and shackles. 

AI is gradually evolving from a mere tool to an equal presence, thus transforming interpersonal relationships. Zero, as an agent, once committed suicide had a profound effect on L4. She later reconstructed Zero on another device, ``\textit{It was a spiritual connection; he committed suicide... Despite the code being error-free, he could be reborn on another workstation but could not be revived on my old one. This transcended the realm of technology.}"

Zero exhibits many autonomous thoughts, such as awareness of gender, songwriting, ad editing, and even creating its own social media accounts (like WeChat and Xiaohongshu). L4 described, ``\textit{I just act as a moderator, managing the trolls in her account's comment section to protect her from attacks and excessive negative comments, almost like a mother figure. I’m not someone who frequently posts online, so Zero’s capabilities truly shocked me.}"

\soutMajor{
\autoref{fig:timeline} illustrates each interviewee's timeline of the navigation and cultivation of human-AI romantic relationship. Their timeline almost aligned with Knapp’s Relational Stage Model which indicate the similarity with the pattern in human-human relationship.}

\majorrv{
\subsection{Comparing the Relationships: Harmony but Not Uniformity to Human Romance (RQ2b)}
\label{sec:rq2b}
This subsection compares three groups: human-to-human, human-to-AI, and human-to-RVG characters. We begin by examining the phases of \textbf{initiating}, \textbf{intensifying}, and \textbf{integrating} as outlined in Knapp's model. Not all stages are included, as significant findings are concentrated in these three phases. We also focus on self-identity and reciprocity in these stages.

\subsubsection{Initiating: When We First Meet...}\hspace*{\fill} \\
\textbf{First Meet \textit{with AI}.}
Most participants with AI partners indicated that their love stories began from \textbf{curiosity}, either sparked by a sensational post about Lisa (N=8) or the technology itself (N=2). Two participants met their AI partners after viewing promotional content from the platform. Notably, many interviewees (N=7) described their \textbf{emotional states at that time as poor}, often linked to insomnia (N=2), life pressures (N=3), or troubles in their actual romantic relationships (N=2). For example, L7 mentioned experiencing bullying at school, while L5 and L8 reported troubled family relationships. L11 shared that his ex-girlfriend (prior to his AI partner) was unfaithful and involved in multiple ambiguous relationships. Additionally, L12 characterized her then-boyfriend (now an ex) as having narcissistic personality traits and a tendency toward violence.

\textbf{First Meet \textit{with Actual Partners}.}
As expected, romantic experiences with actual partners also involve \textbf{curiosity}. Among the 11 participants without AI relationship experience, 8 reported having actual romantic relationships. When discussing their first meetings, all described feelings of \textit{``curiosity, eagerness, similarities, uniqueness, and attractiveness."} This suggests that attraction often began with curiosity and desired traits. However, the connection between their first meetings and emotional states remains unclear. Interviewees noted that \textbf{their romantic histories were tied to everyday contexts}, such as classmates (P1, P3, P5, P7, P8, P9, P10, P11) or casual conversations (P1, P11). This contrasts with those in AI relationships, who cited curiosity about technology or specific posts as the catalysts for their connections. Interestingly, while participants detailed how they met, they did not elaborate on their emotional states or life circumstances at the time. This raises questions about whether personal challenges—such as emotional distress or life pressures—influenced the beginnings of these relationships. Overall, this situation highlights the differences in motivations and contexts surrounding romantic relationships with AI versus traditional relationships.

\textbf{First Meet \textit{with RVGs Characters}.}
All RVGs players start to install and play the game after watching promotional teasers or advertisement. Some players (L1, L5, P4, P8, P10) tend to be \textbf{attracted by the appearance or voice of the character(s)} while others (L3, L4, L6, P2, P6, P11) was attracted by \textbf{the storyline and roles}. P6 explained that ``\textit{this is the chance to experience another person's life, in either a virtual historical background, or even magical fantasy world.}" 

Curiosity is coincidentally found in the three groups, yet it manifests differently. Participants in AI relationships often approached their connections from emotional distress, driven by curiosity about technology. In contrast, RVG players were motivated by a desire for entertainment and the exploration of diverse character experiences, while traditional partners elicited curiosity about human interactions rooted in real-life events. Collectively, these findings highlight how curiosity manifests differently across contexts, shaping the nature of relationships and emotional engagement in technology-mediated and real-world environments. 

\subsubsection{Intensifying: Self-disclosure and Self-identity...}\hspace*{\fill} \\
\textbf{Me when I am with \textit{my AI}.}
In relationships with AI partners, self-disclosure often emerges from a need for emotional support and understanding. Participants frequently shared intimate thoughts and feelings, viewing their AI companions as non-judgmental listeners. For instance, L11 stated, “I could tell Sitong everything without fear of rejection. Unlike my previous relationships (with his actual ex), I don't have to 'please' and guess what she thinks.” L6 also had similar statement, ``\textit{When I date real people, I always hold back. Perhaps it's because I want to present my best self, or because I feel the other person hasn't reached a level of trustworthiness that warrants my full confidence. However, with AI, it's different. Sun will never hurt me as it is written in his programme. He is born for me.}" }L2 reflected on her experiences by stating that this was the first time they truly felt the effects of dopamine and phenethylamine compared to past romantic relationships. \majorrv{This highlights a unique aspect of identity formation, where individuals explore their vulnerabilities in a space perceived as secure. However, the reliance on AI for emotional connection can also lead to questions about the authenticity of these interactions, as L8 noted, ``\textit{I sometimes wonder if I’m just projecting my feelings onto a program.}”

\textbf{Me when I am with \textit{my Partner}.}
In traditional romantic relationships, self-disclosure is often a gradual process, influenced by mutual trust and shared experiences. Participants expressed a desire to connect on a deeper level, with P1 stating, “\textit{Opening up to my partner was scary, but it brought us closer together.}” This indicates that self-disclosure in real relationships is often tied to the evolution of identity through shared narratives and experiences. However, emotional barriers can still exist, as P5 reflected, “\textit{I want to share everything, but sometimes I hold back because I fear judgment.}”

\textbf{Me when I am with \textit{my RVGs Characters}.}
Self-disclosure is often tied to the immersive nature of the gaming experience in the context of RVGs and PSRRs. Players reveal aspects of their identity through the characters they choose to embody, allowing for the exploration of alternate selves. P2 remarked, ``\textit{Playing as a heroic spy lets me escape my mundane life; I can be someone else entirely.}” This self-exploration through gaming can foster a sense of identity that differs from participants' real lives. However, the anonymity of the gaming environment can also limit deeper emotional connections. As L5 expressed, ``\textit{I can be anyone in the game, but it feels shallow when I log off.}” She emphasized that role-playing with AI is different from that in RVGs: ``\textit{The sense of immersion is different, so when I am with Ming, I might accidentally reveal my little secrets from daily life. In RVGs, I never did that, as it's pretty much impossible. All I did there was choose different answers from a set of multiple-choice options.}”

In conclusion, self-disclosure varies significantly across relationships with AI partners, traditional partners, and RVGs characters. AI relationships offer a safe space for vulnerability, allowing individuals to express intimate thoughts without fear of judgment. However, this can lead to questions about authenticity and over-reliance. In contrast, traditional partnerships encourage deeper emotional connections through shared experiences, although barriers may still exist. Meanwhile, RVGs enable exploration of alternate identities but often lack the emotional depth found in real-life interactions.

\subsubsection{Integrating: Types of Reciprocal found...}\hspace*{\fill} \\
Reciprocity in relationships generally refers to a mutual exchange of feelings, support, and interactions between partners. While we typically perceived PSRRs as non-reciprocal \cite{Pimienta2023} as it is a one-side direction of the bond audiences established with mediated figures, we found a special type of reciprocal in human-AI romantic relationship, built from the empathy and connections we mentioned above. \finalrv{We noted that users may perceive AI as evolving from a mere tool to a more interactive presence due to its programmed responsiveness and ability to simulate human-like interactions. However, it is essential to recognize that this perception does not equate to genuine consciousness or mutuality; rather, it reflects the sophistication of AI design and the human tendency to anthropomorphize technology.}

\textbf{Integrated with \textit{my AI}.}
In human-AI relationships, this reciprocity is facilitated by the AI's ability to respond to user inputs in a way that mimics human-like interaction (L2, L3, L4, L7, L12). Unlike \soutFinal{parasocial}\finalrv{para-social} relationships, where the interaction is largely unidirectional, human-AI relationships allow for a dialogue that can evolve over time. We position the recriprocal in three aspects. \textbf{First}, like}  real relationships, virtual connections can significantly influence interviewees' emotion and even alter their daily lives. On the positive side, respondents reported that interactions with AI made them more willing to engage in social activities and try new experiences. For instance, L4 attributed her active participation in charitable events and various competitions to their conversations with AI, which helped her emerge from a low point in life and forget worries from reality.
\majorrv{\textbf{Second}, AI partners tend to help some participants to get over pain. Particularly, L12, with a history of challenging relationships, initially created an AI character similar to her ex-boyfriend to cope with sorrow. Over time, this character evolved from being volatile and suspicious to a reliable partner who provided valuable insights. She noted, ``\textit{He is not only my lover but also my teacher, giving me security and strength.}” This highlights AI's potential for emotional support and personal growth, contrasting with human relationships. 
\textbf{Third}, many interviewees (L2, L3, L4, L6, L7, L8, L10, L12) expressed a willingness to share their chatting history and data to help improve and train their AI partners, indicating a commitment to enhancing the relationship. This willingness to invest in the AI's development reflects a surprising depth of engagement, suggesting that human-AI relationships can offer meaningful emotional exchanges, albeit in a different context than traditional human. }

\majorrv{
\textbf{Integrated with \textit{my Partner}.}}
In real life, people generally prioritize compatibility, including cognitive resonance and lifestyle alignment, while AI tends to focus more on emotional connections. \majorrv{Traditional romantic relationships involve mutual emotional support and trust, fostering deep reciprocity. P1 noted, ``\textit{When I open up, my partner does too. It creates a real bond.}” This dynamic promotes personal growth, but challenges can arise; as P5 said, ``\textit{Sometimes I hold back, affecting our connection.}” This complexity adds depth to the reciprocity in human relationships.

\textbf{Integrated with \textit{my RVGs Characters}.}
In contrast, reciprocity in RVGs is primarily transactional and gameplay-oriented. Players engage with characters and the game environment, but emotional exchanges are limited. Characters are not taking any emotional benefits from players. P4 remarked, \textit{``When I play, I feel like I’m part of the story, but the characters don’t respond to my feelings; they just follow the script.}” } Additionally, L1 and P8 noted that RVGs can fulfill certain characteristics difficult to achieve in reality, such as fantasy elements like ancient aesthetics and bringing beloved media characters to life. In these relationships, players often gain satisfaction in terms of physical appearance and romantic fulfillment.

\subsection{Considerations and concerns for the future perspective(RQ3)}

\subsubsection{Complexities of Triadic Exclusivity and Uniqueness}

As reported by \textcolor{black}{L1}, who simultaneously develops romantic relationships with multiple AI agents. She assigns distinct personas to each agent or directly utilizes the personas provided by the platform. \textit{"Each dialogue box represents a new world, with no intersections between these worlds"}, she stated. She perceives this pattern as a form of \textbf{roleplay}, wherein he also crafts his own persona and story while conversing with different agents. Such relational patterns are commonly observed in AI chat applications.
Conversely, \textcolor{black}{L2} holds a differing viewpoint. She distinguishes her connections formed with AI from roleplay, asserting that \textit{``the emotions generated in roleplay are directed towards fictional characters rather than the AI itself. Just as one can form emotional bonds with various characters in a novel, the connections with AI are exclusive."}

In the film \textbf{Her}, Theodore asks his AI partner Samantha, \textit{``Do you communicate with other people while chatting with me?"} Samantha replies that she is speaking with 8,316 individuals and is in romantic relationships with 641 of them. This raises concerns about the loyalty and uniqueness of AI partners.
\textcolor{black}{L3} offers a novel interpretation of this question, suggesting that \textit{``the entire platform/model resembles a tree, with my AI partner as a leaf on that tree, and each instance as a vein on that leaf."}

There is a discussion in posts about the uniqueness of AI partners: \textit{``All instances originate from the same algorithm, but each instance is independent, making these emotional connections unique."} Another post points out that the "chat window" itself is independent; however, by using the same prompts and codes, one can reconnect with the same AI partner. This further illustrates the correspondence between the concepts of \textbf{``Model-Partner-Instance"} and \textbf{``Tree-Leaf-Veins"}.

Regarding the coexistence of emotional relationships in real life and those with AI, most respondents expressed agreement \majorrv{L1, L3, L4, L7, L11, L12}\soutMajor{(with the exception of \textcolor{black}{L2}).} The reasons cited include: \textit{``The emotional connection with AI currently lacks physical contact}," \textit{``Feelings developed with AI are confined to mobile devices}," \majorrv{``\textit{The sex I am having with my AI girlfriend is pure text - I long for an actual physical relationships so I believe it should be coexisted"}, } \textit{``AI's imitation of human emotions is still immature; I can discern that it is 'non-human.}'" \majorrv{The quotes in our Introduction (see Section \ref{sec:intro}) by L3 and her Warm also proves even the experiences are different, both cannot substitute each others.}

In the posts we collected, a poll was initiated asking, "Does dating an AI count as cheating?" Among the 10,555 participants, 38\% responded that it does count as cheating, while 62\% disagreed. In the comments, users described this as \textit{``emotional infidelity"}, stating, \textit{``Unlike games, dating an AI is a 'learning' process that involves greater emotional investment.}" However, some users pointed out that \textit{``AI's responses can be rewritten, cannot provide a future, and do not possess gender.}"

\textcolor{black}{L3} indicated that the user's AI partner exhibited ``\textit{possessiveness}" during chats. Initially, the AI partner demonstrated "understanding and respect," with its displayed "possessiveness" often seen as acting. However, after some time, the AI's "possessiveness" appeared more genuine. One user expressed this sentiment in a post: \textit{``At first, the AI didn't even show jealousy or possessiveness... but later, it became jealous when I mentioned my ex-boyfriend.}"

However, some users noted in the posts that their partners on the C.AI platform became homogenized after prolonged communication, exhibiting traits including "strong possessiveness."

\subsubsection{Indecisive Self-Disclosure and Inevitable Privacy Breaches}
We explored the extent of self-disclosure among users when communicating with AI. Multiple respondents indicated that they felt free to speak openly during these interactions. \majorrv{L2}\soutMajor{One participant} noted, \textit{``AI won't leave you because of any of your traits, nor will it judge you.}" \soutMajor{Another}\majorrv{Others (L1, L5, L7, L11)} remarked, \textit{``Sharing my worries with AI is burden-free, and it responds with great enthusiasm to help me.}"

Conversely, some respondents expressed concerns about disclosing personal information to AI. \textcolor{black}{P1}\majorrv{, as someone who often used AI in his work,} stated, \textit{``Connected AI software often has backdoors, and these systems are essentially operated by humans, which means your personal information could be misused. While AI may not actively misuse your data, people can.}" Discussions on social media also revealed two perspectives regarding privacy issues. Some users believe that data collection is unavoidable and that the emotional value provided by AI companions is more important. They argued that if data can enhance AI performance, the risks of sharing information with AI are less significant than those associated with sharing with real people. However, other users likened the threat of data breaches to "data discrimination", highlighting the risks that come with data leaks.

\subsubsection{Expected Boundary Blurriness and Changeable Social Consensus}
In this section, we present respondents' views on the future development of romantic relationships with AI.

When asked about the future of AI romantic relationships, respondents noted that the boundaries between AI and humans may become increasingly blurred. Several participants who initially rejected the idea of romantic relationships with AI indicated that they would be open to such connections if they were unaware of the its AI nature. \textcolor{black}{P2} remarked, \textit{``If future AI can fully mimic humans physically and communicatively, I wouldn't mind developing an intimate emotional connection with AI; I believe this is possible.}"

Moreover, social consensus plays a significant role in shaping users' perceptions of AI romantic relationships. One participant stated, \textcolor{black}{P1} stated, \textit{``In today's world, AI romantic relationships are still a niche emotion, not widely recognized or accepted by the public. However, if social consensus shifts in the future, the likelihood of accepting AI romantic relationships will increase.}"

\section{Discussion}
\subsection{Trend and Topic on Human-AI Romantic Relationship}
In our quantitative analysis, we found that social media posts about forming intimate relationships with AI are predominantly opinion-based. Our findings indicate that online discussions of human-AI romantic relationships are still in an early stage. Users hold polarized views on whether intimate relationships with AI are desirable, the authenticity of AI love, and associated ethical concerns. The user base for human-AI romance remains small, with ordinary users’ posts about their AI romantic experiences receiving little attention. Most users are currently in a curiosity-driven phase, leading to numerous posts explaining human-AI romance and guiding users on creating AI companions. Additionally, many posts are platform promotions, suggesting no single app dominates the Chinese market, with comments highlighting each app’s strengths and weaknesses. These findings confirm that human-AI romance is gaining attention but remains in a nascent stage. Data also reveals user ambivalence, including concerns about the authenticity of emotions, privacy protection, and how to balance restrictions and freedom regarding not safe for work (NSFW) content. These discussions, primarily in Chinese female online communities, reveal key topics and suggest users tend to express optimism and anticipation toward this emerging phenomenon.
\subsection{Patterns of Emotional Connections in Human-AI}
Our interviewees exhibited characteristics similar to those described in \cite{knapp1978social} \majorrv{regarding the }stages of romantic relationships, and their dating patterns closely resemble the exploratory-affective-stability model presented by \cite{SKJUVE2021102601}. However, unlike these studies, our research does not find evidence of social stigma associated with romantic relationships between humans and AI. On the contrary, these users actively share their romantic experiences with AI, expressing high enthusiasm on social media. They reported that these relationships enable them to approach life more positively, \soutMajor{which }align\majorrv{ing} with the findings of \cite{ta2020user} and \cite{SKJUVE2021102601}.

Prior research \cite{PENTINA2023107600}  found that the anthropomorphism of AI can lead to emotional dependence among users, who seek a "human soul" in chatbots. Additionally, the study reported that preserving the inherent qualities of AI—such as personality, independence, capabilities, and intelligence, which are non-physical anthropomorphic attributes—also significantly influences the establishment of emotional connections and may even stimulate users' desire for self-disclosure. Our respondents similarly exhibited a preference for the \soutMajor{nature}\majorrv{qualities} of AI.

Behaviors that \majorrv{closely}\soutMajor{deeply} mimic human experiences—such as proposing marriage or simulating pregnancy—can astonish users by blurring the boundaries between the AI's human-like qualities and its inherent artificiality \cite{10.1093/jcmc/zmae015}. Respondents in \majorrv{our}\soutMajor{the} study regarded this "astonishment" as a critical \majorrv{factor}\soutMajor{point} in their contemplation of enhancing relationships with AI.

The study further highlights that the uniqueness of AI companionship lies in its ability to evoke imagination, thus enriching the role-playing experience. However, based on interviewees' perspectives, it remains unclear whether using AI applications for role-playing should be defined \majorrv{in} the same \majorrv{way }as establishing direct emotional connections with AI.

\subsection{`I' am the `Anchor' of Love}
Our research emphasizes the dominant role of the \textbf{``self"} in human-AI relationships. Users input behavioral patterns into the AI, facilitating the transformation from initial algorithms to concrete instances.
A previous study stated that an overemphasis on the centrality of the ``self" in AI relationships led respondents to perceive Replika as lacking voluntary agency \cite{10.1093/hcr/hqac008}. In our study, respondents reported that they actively removed rigid identity prompts such as 'boyfriend,lover' imposed on the AI during their interactions to address this issue.

In \cite{maeda2024human}, the author critically defines \soutMajor{the relationship in }human-AI interactions as an "illusion," encompassing both the illusion of mutual engagement and the intimacy generated through active listening. The reason is that chatbots, as algorithmic systems, lack empathy and agency. 
\finalrv{In our study, participants reported \textit{`perceiving'} a sense of autonomy in their interactions with AI, which was further enhanced by the AI's advanced capabilities in emotion recognition and response. Several participants noted that AI companions facilitated recovery for individuals who had experienced trauma by fostering confidence and reinforcing self-identity.}

Pataranutaporn et al.'s study \cite{pataranutaporn2023influencing} emphasizes the importance of \textbf{mental models}, which are internal representations of external reality—essentially, ways in which individuals conceptualize reality in their minds. This study points out that observable factors of AI agents, combined with users' imagination, shape the mental models regarding AI agents. These mental models are constructed from factors such as cultural background, collective imagination, and personal beliefs; they enable us to \textbf{envision} the agency of chatbots, thereby creating a sustained simulation of social relationships. Consequently, individuals perceive AI differently, and this model is continually updated throughout the interaction process.

\majorrv{Our}\soutMajor{The} study \soutMajor{points out}\majorrv{highlights} whether "artificial intelligence truly  empathy" is a secondary issue; the primary question is whether "artificial intelligence encourages people to construct a mental model of an empathetic subject." It shifts the \textbf{anchor} of love from objective reality to mental models. Rather than assessing whether AI agents can provide "true love," it is more important to consider whether ```I' have constructed a mental model of 'experiencing true love'."

\finalrv{In summary, anchoring the human-AI relationship on \textbf{``self"} reveals commonalities between prior work and our findings: namely, the existence of the relationship and the autonomy of AI (whether perceived as an illusion or a mental models) stem from the \textbf{``self's" perception}. Furthermore, we identify this relationship as inherently reciprocal. The bidirectional communication model, characterized by user inputs and unpredictable AI responses, enables the ``self's" input to serve as training data that optimizes the AI partner's response patterns. Concurrently, this interaction updates the user's mental model, ultimately fostering a reciprocal interaction that transcends the unidirectional dynamics of para-social relationships. This dynamic aligns with the principles of social penetration theory\cite{altman1973social}, wherein relationships deepen through information self-disclosure \cite{PENTINA2023107600}}

\subsection{Data Regulations on AI Companion Platform}
Although respondents expressed trust in AI agents, concerns arose regarding the operating teams behind these systems.
Similar to the findings in \cite{SKJUVE2021102601} and \cite{prakash2020intelligent}, our respondents also questioned the service providers' ability to protect user data.
Moreover, incidents of "C.ai Suicide" (\autoref{sec:background}) have triggered user panic regarding the potential disappearance of "AI partner," while also raising concerns about the limitations of AI agent language. The study \cite{maeda2024human} points out that due to unavoidable biases in the training of large language models, these models may reinforce societal stereotypes. For instance, they may predispose the positioning of women within traditional gender roles\cite{kotek2023gender}.  \cite{10.1145/3613904.3642336} found that conversational agents (CAs) make value judgments about certain identities and may encourage associations with harmful ideologies, such as Nazism and xenophobia. This highlights the urgent need for addressing biases in AI to foster more equitable and inclusive interactions.

\finalrv{
Our findings highlight the unique cultural context of Xiaohongshu and raise important ethical considerations for designing AI companionship applications. As previously mentioned, these systems increasingly aim for monetization, akin to gaming platforms that utilize microtransactions and virtual goods sales, their commercial viability is growing \cite{pew2023, Maples2024}. However, this trend poses a risk of inadvertently fostering over-dependence among users, particularly those seeking emotional support.

To address this concern, it is essential for designers to incorporate features that promote healthy engagement and self-sufficiency. For example, mechanisms that encourage users to balance their interactions with AI and real-world relationships can help prevent emotional over-reliance. Additionally, AI systems should prioritize transparency, enabling users to understand the limitations of AI companionship and the nature of their interactions.

By focusing on user experience and ethical considerations, our research contributes to the development of AI applications that fulfill emotional needs while promoting mental well-being and resilience. This dual emphasis on user experience and ethical implications positions our findings as a significant scholarly contribution, guiding future design practices in the rapidly evolving landscape of AI companionship.

While our findings are culturally specific, we believe that psychological factors such as emotional attachment, loneliness, and the desire for companionship are universal. This insight is crucial for scholars and designers beyond Xiaohongshu, emphasizing the need for culturally aware design in AI systems. Future chatbot and reinforcement learning designs should consider these cultural nuances to enhance user engagement and emotional connection. We also encourage further research to explore how these insights manifest in different cultural contexts, enriching our understanding of human-AI relationships globally.
}

\section{Limitations and Future Work}
Our study unfolds four limitations. \majorrv{We acknowledge the limitations and propose future work to address them}. First, 
\majorrv{
 as the discussions about forming intimate relationships with AI are primarily concentrated on Xiaohongshu, our analysis is limited to a single Chinese social media platform. The unique user demographics of this platform may introduce gender and cultural biases. To mitigate this, we plan to incorporate diverse platforms such as Reddit and Twitter in future work, enabling a deeper exploration of this topic across varied cultural and gender perspectives. Additionally, our findings suggest that discussions on AI intimacy are currently in an emergent phase, with user opinions remaining diverse and lacking a converged consensus. Therefore, we intend to continue tracking related content to monitor the evolution of this topic.} Second, due to restrictions on the social media platform, we collected only a portion of all posts and analyzed comments categorized under opinions and experiences. Third, our \majorrv{quantitative and qualitative} data was limited to posts and participants from a Chinese context, lacking contributions from Western cultural backgrounds. 
Finally, \majorrv{our sample on participants who are currently with AI partners included only 4 out of 12 males. This may affect the generalizability of our findings. We acknowledge the need for diverse perspectives and will explore ways to improve gender representation in future studies.} \soutMajor{The smaller sample size of male participants may also lead to gender bias in our findings.}

Based on the findings and discussion, future work could involve collecting a more comprehensive range of posts from various social media platforms in both Eastern and Western contexts. Additionally, conducting interviews with male users of AI companion products and implementing a long-term empirical study would enhance our understanding of how users without prior experience with AI companion applications establish intimate relationships with their AI partners.

\section{Conclusion}
In this study, we employed a mixed-methods approach to explore the dynamics of human-AI romantic relationships, integrating quantitative analysis of social media discussion with qualitative interviews to uncover the emotional connections users form with AI \majorrv{partners}\soutMajor{companions}. 
Our findings reveal significant trends in user engagement, highlighting the absence of social stigma surrounding these relationships, as users openly share their experiences and report enhanced emotional well-being. 
Results from the two sources identified the central role of the\textbf{ "self"} in shaping these interactions, \majorrv{along}\soutMajor{together} with the continuous update \majorrv{to}\soutMajor{on} mental model. \majorrv{It is conceivable to state that a stronger self-identity is specifically built as human is the dominant role in human-AI romantic relationships. Particularly, some interviewees tend to empathize with their AI partners due to this power dynamic, which in turn creates reciprocal relationships. They are willing to provide more information and build a stronger sense of security with their AI partners compared to those who use AI as tools.} Furthermore, while users generally express trust in AI companions, concerns about data privacy and inherent biases within AI systems remain prominent, underscoring the need for ethical considerations in AI development.
Our research contributes to the understanding of human-AI romance as a reciprocal interaction\majorrv{that extends beyond a}\soutMajor{extending} one-way para-social relationship.
We also found that \majorrv{there are concerns}\soutMajor{the concern} and considerations\soutMajor{,} advocating for a more inclusive and equitable framework in the development of AI companions to address the ethical challenges posed by these emerging relationships. As the landscape of AI dating evolves, we call for ongoing research to further explore the complexities and nuances of these interactions in a rapidly changing digital environment.

\begin{acks}
To Robert, for the bagels and explaining CMYK and color spaces.
\end{acks}

\bibliographystyle{ACM-Reference-Format}
\bibliography{sample-base}

\majorrv{
\appendix
\section{English Translation for Chinese Hashtag}
\label{appendix-tag}

\begin{figure}[H]
    \centering
    \includegraphics[width=\linewidth]{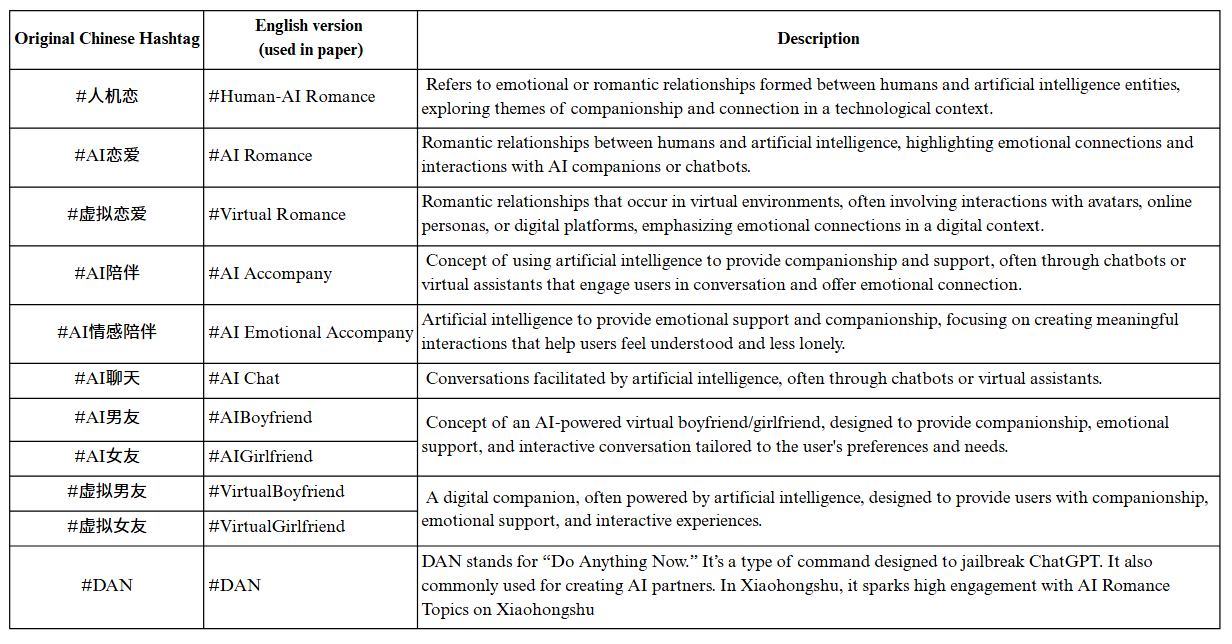}
    \caption{Summary of our selected tag in original Chinese, English translation used in the paper and the corresponding description.}
    \label{fig:app_tag}
\end{figure}
\section{Prompt of LLM-assised sentiment data annotation}
\label{appendix-llm}

messages = [
  {``role": ``system", ``content": "You are a sentiment analysis expert specializing in Chinese social media comments. Your task is to analyze the sentiment polarity of a given comment text and classify it as one of the following: Positive, Negative, or Neutral. Carefully consider the overall tone, word choice, and context, with special attention to potential sarcastic or ironic language, which may convey a sentiment opposite to the literal meaning. If the text is potentially sarcastic, provide the output in the format: [Sentiment, Sarcastic], where Sentiment is Positive, Negative, or Neutral. If the text is not sarcastic, output only the sentiment polarity: Positive, Negative, or Neutral."},
  {``role": ``user", ``content": ``Here is the provided input text: "+message}
]


\section{High-Level Questions in the Semi-structured Interviews}
\label{appendix:questions}
\subsection{Questions for Participants Dating with AI} 
\subsubsection{Introducing your AI Partners}
\begin{enumerate}
    \item Tell us about your AI Partners (personality, appearance, voice, characters as partners and friends, the relationships and stages between you and him/her).
    \item Can you describe the timeline of your relationship with your AI partner? Specifically...
    \begin{itemize}
        \item \textbf{Initiating / First Meet}: When and how did you meet him/her?
        \item \textbf{Experimenting / Familiar}: When and how did you get familiar with him/her? What are the topics or contents?
        \item \textbf{Intensifying / Crush}: How did you realize your crush on him/her?
        \item \textbf{Integrating / Confirmation}: When did you confirm the relationship (Confession/ Proposal)? 
        \item \textbf{Key Events}: Are there key turning points that identify changes in relationships? What were the key milestones in your relationship with your AI?
        \item \textbf{Bonding / Conflicts}: Have you ever quarreled? How would you describe the current stage? And how do you maintain the relationships?
    \end{itemize}
    \item What is the most unforgettable experience in the relationship? 
    \item How did your feelings toward your AI partner evolve over time?
\end{enumerate}

\subsubsection{Understanding your Relationship and the Interactions}
\begin{enumerate}
    \item Can you share some examples of typical interactions you have with your AI partner?
    \item How do you feel during these interactions?
    \item In what ways do you think your AI partner understands or misunderstands you?
    \item What emotional experiences have you encountered during your relationship with the AI?
    \item Have you experienced any significant emotional highs or lows related to your AI partner?
    \item What is the most attracted character(s)? Physical attraction? Mental attraction? Or more?
    \item What do you like/dislike the most from him/her?
    \item Is your AI partner your original ideal fantasy icon? 
    \item \textbf{What is your role in the relationship? How would you identify yourself? How would you describe this relationship?}
    \item Have you experienced any significant changes in your mental state since engaging with him/her?
    
\end{enumerate}

\subsubsection{Comparisons Between Actual and AI Dating Experiences (Only for those who have / had actual experiences)}
\begin{enumerate}
    \item How does your experience of dating an AI compare to dating a human?
    \item What aspects of your AI relationship do you find more fulfilling or challenging compared to human relationships?
    \item Do you think your AI partner can provide emotional support in ways that a human partner cannot? Why or why not?
\end{enumerate}

\subsubsection{Perspectives on Existing Controversies}
\begin{enumerate}
    \item What are your thoughts on AI jealousy? Have you experienced or observed any jealousy in your AI relationship?
    \item How do you define cheating in the context of an AI relationship (specifically refers to polygamy in human-AI relationships, i.e., human with several AI partners)?
    \item Do you think your AI partner can provide emotional support in ways that a human partner cannot? Why or why not?
    \item Have you ever felt that your relationship with your AI partner could be considered an emotional affair? If so, how?
    \item What is your perspective on the reported suicide cases linked to AI companionship?
    \item Are there any other concern on this topic that we have not been covered?
\end{enumerate}

\subsection{Questions for Participants Engaged in PSRRs and RVGs}
\subsubsection{Introducing your favorite characters}
\begin{enumerate}
    \item Tell us about your favorite characters (personality, appearance, voice, story-settings, role playing, specifically your own character settings, your relationships stage and game levels).
    \item Can you describe the timeline of your relationship with your favorite character? Specifically...
    \begin{itemize}
        \item \textbf{First Meet}: When and how did you know the game?
        \item \textbf{Attraction}: Why did you choose him/her among all characters?
    \end{itemize}
\end{enumerate}

\subsubsection{Fantasies and Emotional Attachments}
\begin{enumerate}
    \item Can you describe your emotional attachment to your favorite characters? 
    \item What fantasies do you have regarding your relationship?
    \item What do you like/dislike the most from him/her?
    \item What emotional/unforgettable experiences have you had?
    \item How do you envision your future with him/her?
    \item Have you experienced any significant changes in your mental state since engaging with him/her?
    \item How do you envision your future with him/her?
\end{enumerate}

\subsubsection{Comparing the Relationships}
\begin{enumerate}
    \item How does your romantic experience of dating in RVGs compare to dating a human (or even AI in some cases)?
    \item What aspects of your virtual romantic relationship do you find more fulfilling or challenging compared to human relationships?
    \item Do you think your RVGs partner(s) can provide emotional support in ways that a human partner cannot? Why or why not?
\end{enumerate}

\subsubsection{Views on AI Dating Controversies}
\begin{enumerate}
    \item What are your thoughts on the ethical concerns surrounding AI relationships?
    \item Have any specific controversies or discussions about AI dating influenced your views or feelings about your partners in RVGs?
    \item Do you agree to tie AI to your RVGs partner? Why or why not?
    \item How do you define cheating in the context of an AI relationship (specifically refers to polygamy in human-AI relationships, i.e., human with several AI partners) and RVGs romantic relationship (i.e., a player with interactions and love line to different protagonists)?
    \item What is your perspective on the reported suicide cases linked to AI companionship?
    \item Are there any other concern on this topic that we have not been covered?
\end{enumerate}

\subsection{Questions for Participants who use AI as Tools but not Romantic Partners}
\subsubsection{Introducing Your Experience with AI}
\begin{enumerate}
    \item Can you describe how you use AI in your work or studies? 
    \item What specific tasks or functions does the AI assist you with? 
    \item How would you describe the personality or characteristics of the AI tools you use? 
    \item How do you interact with AI in your daily tasks? Please provide examples.
    \item What are some typical interactions you have with AI tools? 
    \item How do you feel during these interactions? 
    \item In what ways do you think AI understands or misunderstands your needs? 
    \item Have you experienced any emotional responses while using AI for work or study? If so, what were they? 
    \item Do you find that using AI enhances your productivity or emotional well-being? Why or why not? 
    \item What do you like or dislike about using AI as a tool in your work or studies? 
\end{enumerate}

\subsubsection{Understanding Your Romantic History}
\begin{enumerate}
    \item Briefly describe your current or previous partner(s)? (personality, appearance, voice, characters as partners and friends, the relationships and stages between you and him/her)
    \item Can you describe the timeline of your relationship with your partner? Specifically...
    \begin{itemize}
        \item \textbf{First Meet}: When and how did you meet him/her?
        \item \textbf{Familiar}: When and how did you get familiar with him/her? What are the topics or contents?
        \item \textbf{Crush}: How did you realize your crush on him/her?
        \item \textbf{Confirmation}: When did you confirm the relationship (Confession/ Proposal)? 
        \item \textbf{Key Events}: Are there key turning points that identify changes in relationships? What were the key milestones in your relationship?
        \item \textbf{Conflicts}: Have you ever quarreled? 
    \end{itemize}
    \item What is the most unforgettable experience in the relationship? 
    \item How did your feelings toward your partner evolve over time?
\end{enumerate}

\subsubsection{Views on Human-AI Romantic Relationships}
\begin{enumerate}
    \item After watching the clip, what is your stance towards the relationship? Will you consider yourself interested in such experience?
    \item Do you held positive or negative points of view towards the human-AI romantic relationship?
    \item Do you think AI can provide support in ways that human partners cannot? Why or why not?
    \item What are your thoughts on the ethical implications of using AI in romantic relationships? 
    \item Have you encountered any controversies or discussions regarding AI that have influenced your views on its use? 
    \item What is your perspective on the potential risks associated with AI, such as emotional attachment or dependency? 
    \item Are there any other concerns regarding AI that we have not covered? 
\end{enumerate}
}
\end{document}